\documentclass[fleqn,usenatbib]{mnras}

\usepackage[T1]{fontenc}
\usepackage{ae,aecompl}

\usepackage{graphicx}        
\usepackage{amsmath}         
\usepackage{amssymb}         

\usepackage{booktabs}        
\usepackage{scalerel}        
\usepackage{xcolor,listings} 
\usepackage{upquote}         
\usepackage{multirow}        
\usepackage{pdflscape}       
\usepackage{placeins}

\newcommand*{\dirplot}{plot/png_300}
\newcommand*{\ext}{png}

\newcommand*{\pd}[1]{\!\times\!10^{#1}}
\newcommand*{\units}[1]{\scalebox{0.8}{(#1)}}

\newcommand*{\Sun}{\protect\scalebox{0.7}{$\odot$}}
\newcommand*{\Star}{{\rm bar}}

\newcommand*{\Range}[2]{[#1\,\text{,}\,#2]}

\newcommand*{\diff}{\mathop{}\!\mathrm{d}}

\newcommand*{\SM}[1]{{\scaleto{\rm #1}{3.5pt}}}

\newcommand*{\var}[1]{\mbox{\footnotesize$(#1)$}}
\newcommand*{\vvar}[1]{\mbox{\footnotesize$\big(#1\big)$}}

\newcommand*{\plus}{\scalebox{0.75}[1.0]{$+$}}

\newcommand*{\Gband}{\mbox{$G\:\!\text{-band}$} }

\newcommand*{\GtrSim}{\smallrel\gtrsim}
\newcommand*{\LessSim}{\smallrel\lesssim}

\newcommand*{\Approx}{\smallrel\sim}

\makeatletter
\newcommand*{\smallrel}[2][.8]{%
  \mathrel{\mathpalette{\smallrel@{#1}}{#2}}%
}
\newcommand*{\smallrel@}[3]{%
  \sbox0{$#2\vcenter{}$}%
  \dimen@=\ht0 %
  \raise\dimen@\hbox{%
    \scalebox{#1}{%
      \raise-\dimen@\hbox{$#2#3\m@th$}%
    }%
  }%
}

\DeclareMathOperator{\logN}{logN}

\definecolor{myred}{HTML}{D62728}

\newcommand*{\MW}{\protect\scalebox{0.65}{MW}}
\newcommand*{\MC}{\protect\scalebox{0.65}{MC}}

\title[The oblateness of the Milky Way halo]{
The oblateness of the Milky Way dark matter halo from the stellar streams of NGC 3201, M68, and Palomar 5}

\author[C. G. Palau, J. Miralda-Escud\'e]{
Carles G. Palau $^{1}$\thanks{E-mail: cgpalau@sjtu.edu.cn},
Jordi Miralda-Escud\'e $^{1,2}$\thanks{E-mail: miralda@icc.ub.edu}
\\
$^{1}$Institut de Ci\`encies del Cosmos, Universitat de Barcelona (UB-IEEC), Mart\'i i Franqu\`es 1, E-08028 Barcelona, Catalonia, Spain.\\
$^{2}$Instituci\'o Catalana de Recerca i Estudis Avan\c cats, E-08028 Barcelona, Catalonia, Spain.\\
}

\date{Accepted XXX. Received YYY; in original form ZZZ}

\pubyear{2023}

\begin{document}
\label{firstpage}
\pagerange{\pageref{firstpage}--\pageref{lastpage}}
\maketitle

\begin{abstract}
We explore constraints on the Milky Way dark matter halo oblateness using three stellar streams from globular clusters NGC 3201, M68, and Palomar 5. Previous constraints on the gravitational potential from dynamical equilibrium of stellar populations and distant Milky Way satellites are included. We model the dark halo as axisymmetric with axis ratio $q_\rho^{\rm h}$ and four additional free parameters of a two power-law density profile. The halo axis ratio, while barely constrained by the NGC 3201 stream alone, is required to be close to spherical by the streams of Palomar 5 ($q_\rho^{\rm h}=1.01\pm0.09$) and M68 ($q_\rho^{\rm h}=1.14^{+0.21}_{-0.14}$), the latter allowing a more prolate shape. The three streams together are well fitted with a halo axis ratio $q_\rho^{\rm h}=1.06\pm0.06$ and core radius $\Approx$ 20 kpc. Our estimate of the halo shape agrees with previous studies using other observational data and is in tension with cosmological simulations predicting that most spiral galaxies have oblate dark matter halos with the short axis perpendicular to the disc. We discuss why the impact of the Magellanic Clouds tide is too small to change our conclusion on the halo axis ratio. We note that dynamical equilibrium of a spherical halo in the oblate disk potential implies an anisotropic dark matter velocity dispersion, larger along the vertical direction than the horizontal ones, which should relate to the assembly history of the Milky Way.
\end{abstract}




\defcitealias{2019MNRAS.488.1535P}{PM19}
\defcitealias{2021MNRAS.504.2727P}{PM21}

\section{Introduction}

The dark halo of the Milky Way is the least known component of our Galaxy.
Determining its density profile and three-dimensional shape is an important
astrophysical goal that can help us understand how galaxies form and evolve
and constrain the properties of the dark matter.

Simulations of the formation and evolution of galaxies have been one of
the main tools to predict the shape of the dark halo of galaxies similar
to the Milky Way \citep{2020NatRP...2...42V}. In general, simulations including only dark matter
produce halos with triaxial shapes following the Navarro, Frenk \& White
\citep[NFW,][]{1996ApJ...462..563N} density profile. When baryons are
included, interactions between baryons and dark matter in disc galaxies
make halos rounder and approximately axisymmetric, with the minor axis
perpendicular to the disc \citep[e.g.][]{2005ApJ...627L..17B,2012MNRAS.426..983D,2020PDU....2800503D}.

Testing these predictions from observations has proved difficult \citep[see e.g.][]{2018ARAandA..56..435W, 2019AandARv..27....2S}.
Galaxy rotation curves provide ambiguous constraints on the shape of the
dark halo because of the uncertainties in subtracting the baryonic
component of stars and gas, and depend only on the potential in the disc
plane \citep{2020ApJS..247...31L}. In the Milky Way, dynamical equilibrium methods of tracers like
globular clusters or halo stars, as well as the orbits from stellar debris
of the Sagittarius dwarf galaxy, have been applied to constrain the
potential \citep[e.g.][]{2019MNRAS.483.4724F,2019MNRAS.485.3296W,2021MNRAS.508.5468H}. The shape of the dark halo is still poorly constrained by
these methods, and varying results of oblate, prolate, spherical, and
triaxial configurations have been obtained depending on the method and
the source of observational data (see Section \ref{CTOS}).

Here we use dynamically cold stellar streams to study the shape of the Milky Way's halo. These structures are formed when
a progenitor satellite galaxy or globular cluster is perturbed by tidal
shocks, generally when the progenitor approaches the centre of the
galaxy o crosses the disc \citep[e.g.][]{2008MNRAS.387.1248K,2012MNRAS.420.2700K,2014ApJ...795...95B}. The ensuing loss of stars from the bound
system populates the leading and trailing tails of the stream. The
tidally stripped stars approximately follow the orbit of the progenitor
with a small variation of the orbital energy, with stars that gain
energy moving to the trailing arm (a longer period orbit), and those
that lose energy moving to the leading arm (a shorter period orbit).
Models of the phase-space structure of stellar streams can help
reconstruct the orbit of the progenitor and use it to constrain the
gravitational potential of the galaxy
\citep[see e.g.][]{2011MNRAS.417..198V, 2014ApJ...794....4P, 2014ApJ...795...94B}.

Several streams have been discovered in the inner region of the Milky
Way \citep[e.g.][]{2006ApJ...637L..29B, 2006ApJ...639L..17G,2018ApJ...862..114S}, and some of them have been used to constrain the potential of the Galaxy (see Section \ref{CTOS}). For example, the GD-1 stellar stream, one of the most populated, has no known progenitor. This makes it difficult to model and constrain the Galactic potential with this stream. Another prominent stellar stream is the one generated by the Palomar 5 globular cluster, at $\Approx$ 16 kpc from the Galactic centre and far above the disc. This location is ideal to study the inner halo shape because the trajectory of the stream depends on the vertical acceleration, which is sensitive to the halo oblateness \citep[see e.g.][]{2015ApJ...799...28P}.

The publication of the second version of the \textit{Gaia} star
catalogue (GDR2), with more than 1 billion sources
\citep{2016AandA...595A...1G,2018AandA...616A...1G}, has improved the quality of the existing data of the Palomar 5 tidal stream, providing parallaxes and proper motions of many stars along the stream. Furthermore, this catalogue has made it possible to discover other stellar streams \citep[see e.g.][]{2018ApJ...865...85I,2018MNRAS.481.3442M,2019ApJ...872..152I}, some of them associated with globular clusters \citep{2019ApJ...884..174G,2019NatAs...3..667I,2020AandA...637L...2P}. Two of the main examples are the streams of M68 and NGC 3201 \citep[][hereafter PM19, PM21]{2019MNRAS.488.1535P,2021MNRAS.504.2727P}.
These streams are dynamically cold and relatively close to the Sun,
greatly facilitating their study with the \textit{Gaia} data.
Each stellar stream provides independent constraints on the Milky Way
mass distribution, helping resolve degeneracies that inevitably arise
when modeling all the Milky Way components with many parameters.

 In this paper, we present a method to constrain a model of the Milky Way halo
using several stellar streams combined with other traditional observational
constraints. We apply it to the streams of NGC 3201, M68, and
Palomar 5. This combination of multiple
observations is essential to help separate the contributions from the
disc, bulge and halo, and reduce model degeneracies.
In Section \ref{MassModel}, we discuss our mass model of the Galaxy and the prior constraints on the free parameters from observational data. In Section \ref{KCons}, we present the kinematic constraints and a description of each stellar stream. In Section \ref{SM}, the stream-fitting methodology is explained and the method is applied to the observational data. Results with each stream separately are presented in Section \ref{Res} and for all streams together in Section \ref{AllS}. In Section \ref{CTOS} we compare the halo axis ratio to previous estimates in the literature, and we present our conclusions in Section \ref{con}.

\section{Mass model of the Milky Way}\label{MassModel}

We model the mass distribution of the Milky Way as the sum of three components: disc, bulge and halo. We now describe the parameterized models
used for each of them.

\subsection{The disc density profile}

The Milky Way stellar disc is modelled as the sum of two exponential
profiles for the thin and thick disc. We do not separate the contribution of gas from stars; the total gas mass is approximately $M_{\rm gas} \Approx 10^{10}$ M$_{\Sun}$, smaller than the stellar mass of $M_{\rm d}\Approx 4\pd{10}$ M$_{\Sun}$ \citep[e.g.][]{2016ARAandA..54..529B}, and we neglect the different scale heights for the gas and stellar components. We note that the thin gas and young stars component may increase the strength of tidal shocks when crossing the disc and therefore the number and ejection velocities of stars that populate the tidal tails, so a more precise modeling of the vertical profile will be useful in future work.

In Galactocentric Cylindrical coordinates $(R,\varphi,z)$, the mass density for each stellar component is
\begin{equation}
\rho_\gamma\var{R,z} = \frac{\Sigma_\gamma}{2z_\gamma} 
 \exp\left( -\frac{R}{h_\gamma} -\frac{|z|}{z_\gamma} \right) ~,
\end{equation}
where the subindex takes two values: $\gamma={\rm n}$ denotes the thin disc,
and $\gamma={\rm k}$ the thick disc. The central mass surface density is
$\Sigma_\gamma$, $h_\gamma$ is the radial scale length, and $z_\gamma$
the vertical scale height. The scale lengths and scale heights are
constrained at the solar vicinity by star counts in optical and infrared
bands to values $h_{\rm n} \Approx 2.5$ kpc, $z_{\rm n} \Approx 300$ pc
for the thin disc, and $h_{\rm k} \Approx 2$ kpc,
$z_{\rm k} \Approx 900$ pc for the chemically defined thick disc
\citep[e.g.][]{2008ApJ...673..864J,2015ApJ...800...83B}. The mass surface
density ratio of the two components is also estimated in the solar
vicinity \citep[e.g.][]{2008ApJ...673..864J,2010MNRAS.402..461J}.

As a consistent methodology to fit our mass distribution model to
various observations, we will let these model parameters vary in our maximum a-posteriori fits. These parameters are also constrained by Gaussian priors defined by various observational determinations with estimated errors. We choose the estimates for scale lengths and scale heights given in the review article of \citet{2016ARAandA..54..529B}, and we list them in Table \ref{FreePar} with their errors that are assumed to be uncorrelated. The surface densities $\Sigma_k$ and $\Sigma_n$ are left free with a uniform positive prior. We add to the likelihood function (see Section \ref{SM}) the constraint on the local ratio of the thin and thick disc surface densities
\begin{equation}\label{surf_den_ratio}
f_\varSigma \equiv f_\rho \frac{z_{\rm k}}{z_{\rm n}} = 0.12 \pm 0.04 ~,
\end{equation} 
where $f_\rho \equiv \rho_{\rm k}\var{R_{\Sun}, z_{\Sun}} / \rho_{\rm n}\var{R_{\Sun}, z_{\Sun}} $ is the local density ratio. We also take this measurement from \citet{2016ARAandA..54..529B}.

In general, Table \ref{FreePar} lists all our variable parameters, with indication of their priors, and Table \ref{FixPar} lists all our fixed parameters, for which we consider their errors to be of negligible impact for our modeling purpose.

\begin{table}
\caption[]{\small{Free parameters $\theta$. The priors $p_\theta$ are assumed to be Gaussian distributions $\mu\pm\sigma$ with mean $\mu$ and standard deviation $\sigma$ or uniform distributions when they are not specified.}}
\begin{center}
\begin{tabular}{lllc}
\toprule
\multicolumn{2}{l}{\textbf{Sun}}&\textbf{Gaussian Prior}&\textbf{Ref.}\\
\midrule
$R_{\Sun}$&\units{kpc}&$8.178\pm0.026$&[1]\\
$U_{\Sun}$&\units{km s$^{-1}$}&$11.1\pm1.25$&[2]\\
$V_{\Sun}$&\units{km s$^{-1}$}&$12.24\pm2.05$&[2]\\
$W_{\Sun}$&\units{km s$^{-1}$}&$7.25\pm0.62$&[2]\\
\midrule
\multicolumn{2}{l}{\textbf{Disc}}&&\\
\midrule
$\varSigma_{\rm n}$&\units{${\rm M}_{\Sun}\,{\rm kpc}^{-2}$}&&\\
$h_{\rm n}$&\units{kpc}&$2.6\pm0.5$&[3]\\
$z_{\rm n}$&\units{kpc}&$0.3\pm0.05$&[3]\\
\midrule
$\varSigma_{\rm k}$&\units{${\rm M}_{\Sun}\,{\rm kpc}^{-2}$}&&\\
$h_{\rm k}$&\units{kpc}&$2.0\pm0.2$&[3]\\
$z_{\rm k}$&\units{kpc}&$0.9\pm0.18$&[3]\\
\midrule
\multicolumn{2}{l}{\textbf{Bulge}}&&\\
\midrule
$\rho_{0}^{\rm b}$&\units{${\rm M}_{\Sun}\,{\rm kpc}^{-3}$}&&\\
\midrule
\multicolumn{2}{l}{\textbf{Dark halo}}&&\\
\midrule
$\rho_{0}^{\rm h}$&\units{${\rm M}_{\Sun}\,{\rm kpc}^{-3}$}&&\\
$\alpha$&&&\\
$a_{1}$&\units{kpc}&&\\
$\beta$&&&\\
$q_{\rho}^{\rm h}$&&&\\
\midrule
\multicolumn{2}{l}{\textbf{NGC 3201}}&&\\
\midrule
$r_{\rm h}$&\units{kpc}&$4.9\pm0.11$&[4]\\
$v_r$&\units{km s$^{-1}$}&$494.34\pm0.14$&[5]\\
$\mu_\delta$&\units{mas yr$^{-1}$}&$-1.991\pm0.044$&[6]\\
$\mu_{\alpha*}$&\units{mas yr$^{-1}$}&$8.324\pm0.044$&[6]\\
\midrule
\multicolumn{2}{l}{\textbf{M68 (NGC 4590)}}&&\\
\midrule
$r_{\rm h}$&\units{kpc}&$10.3\pm0.52$&[4]\\
$v_r$&\units{km s$^{-1}$}&$-92.99\pm0.22$&[5]\\
$\mu_\delta$&\units{mas yr$^{-1}$}&$1.762\pm0.053$&[6]\\
$\mu_{\alpha*}$&\units{mas yr$^{-1}$}&$-2.752\pm0.054$&[6]\\
\midrule
\multicolumn{2}{l}{\textbf{Palomar 5}}&&\\
\midrule
$r_{\rm h}$&\units{kpc}&$20.6\pm0.2$&[7]\\
$v_r$&\units{km s$^{-1}$}&$-58.6\pm0.21$&[5]\\
$\mu_\delta$&\units{mas yr$^{-1}$}&$-2.646\pm0.064$&[6]\\
$\mu_{\alpha*}$&\units{mas yr$^{-1}$}&$-2.736\pm0.064$&[6]\\
\bottomrule
\end{tabular}
\end{center}

\begin{tabular}{l}
\textit{Note.}\\
\text{[1]}: \citet{2019AandA...625L..10G}\\
\text{[2]}: \citet{2010MNRAS.403.1829S}\\ 
\text{[3]}: \citet{2016ARAandA..54..529B}\\
\text{[4]}: \citet{1996AJ....112.1487H, 2010arXiv1012.3224H}\\
\text{[5]}: \citet{2019MNRAS.482.5138B}\\
\text{[6]}: \citet{2019MNRAS.484.2832V}\\
\text{[7]}: \citet{2019AJ....158..223P}\\
\end{tabular}

\label{FreePar}
\end{table}

\subsection{The bulge density profile}

 We consider the Milky Way bulge and bar \citep[see e.g.][]
{2015MNRAS.448..713P, 2015MNRAS.450.4050W, 2019MNRAS.489.3519C} as a single component in this paper. In our case, the streams we are studying do not penetrate to the innermost part of the Galaxy and their dynamics are therefore only weakly affected by the detailed mass distribution of this component. We assume for simplicity an axisymmetric bulge with a power-law density profile with core $h_{\rm b}$, slope $\alpha_{\rm b}$ and a Gaussian truncation at a scale length $a_{1\rm b}$,
\begin{equation}
\rho_{\rm b}\var{s} = \rho_{0}^{\rm b} \left(1+\frac{s}{h_{\rm b}}\right)^{-\alpha_{\rm b}} \exp\!\left(-\frac{s^2}{a_{1\rm b}^2} \right) ~,
\end{equation}
which is constant over ellipsoids of constant $s$,
\begin{equation}\label{s}
s^2 \equiv R^2 + \frac{z^2}{q_\rho^2} ~,
\end{equation}
with axis ratio $q_{\rho}=q_{\rho}^{\rm b}$. This model is an axisymmetric version of \cite{2002MNRAS.330..591B} introduced by \citet{2011MNRAS.414.2446M}. We fix all the bulge parameters following \citet{2017MNRAS.465...76M} to the values listed in Table \ref{FixPar}, except for the density normalization parameter $\rho_{0}^{\rm b}$, which we leave as a free parameter. We note that we have not imposed any central hole in the surface density model of the disc, so our model for the central bulge is a rough one because the resulting mass distribution includes the central part of our exponential disc. The scale density is proportional to the bulge mass $M_{\rm b}$, which we constrain in the range following \citet{2017MNRAS.465...76M}:
\begin{equation}
M_{\rm b} = (8.9 \pm 0.89)\pd{9} \,\, {\rm M}_{\Sun} ~.
\end{equation}

\begin{table}
\caption[]{\small{Fixed properties of the Sun, bulge, and globular clusters.}}
\begin{center}
\begin{tabular}{lllc}
\toprule
\multicolumn{2}{l}{\textbf{Sun}}&\textbf{Value}&\textbf{Ref.}\\
\midrule
$z_{\Sun}$&\units{pc}&$25$&[1]\\
\midrule
\multicolumn{2}{l}{\textbf{Bulge}}&&\\
\midrule
$h_{\rm b}$&\units{pc}&$75$&[2]\\
$a_{1 \rm b}$&\units{kpc}&$2.1$&[2]\\
$q_{\rho}^{\rm b}$&&$0.5$&[2]\\
$\alpha_{\rm b}$&&$1.8$&[2]\\
\midrule
\multicolumn{2}{l}{\textbf{NGC 3201}}&&\\
\midrule
$M_{\rm gc}$&\units{$10^4\,{\rm M}_{\Sun}$}&$6.47$&[5]\\
$a_{\rm gc}$&\units{pc}&$4.9$&[5]\\
$\delta$&\units{deg}&$-46.412$&[4]\\
$\alpha$&\units{deg}&$154.403$&[4]\\
\midrule
\multicolumn{2}{l}{\textbf{M68 (NGC 4590)}}&&\\
\midrule
$M_{\rm gc}$&\units{$10^4\,{\rm M}_{\Sun}$}&$5.7$&[3]\\
$a_{\rm gc}$&\units{pc}&$6.4$&[3]\\
$\delta$&\units{deg}&$-26.744$&[4]\\
$\alpha$&\units{deg}&$189.867$&[4]\\
\midrule
\multicolumn{2}{l}{\textbf{Palomar 5}}&&\\
\midrule
$M_{\rm gc}$&\units{$10^3\,{\rm M}_{\Sun}$}&$4.3$&[5]\\
$a_{\rm gc}$&\units{pc}&$8.43$&[5]\\
$\delta$&\units{deg}&$-0.112$&[4]\\
$\alpha$&\units{deg}&$229.022$&[4]\\
\bottomrule
\end{tabular}
\end{center}

\begin{tabular}{l}
\textit{Note.}\\
\text{[1]}: \citet{2008ApJ...673..864J}\\
\text{[2]}: \citet{2017MNRAS.465...76M}\\
\text{[3]}: \citet{2010MNRAS.406.2732L}\\
\text{[4]}: \citet{1996AJ....112.1487H, 2010arXiv1012.3224H}\\
\text{[5]}: \citet{2017MNRAS.471.3668S}\\
\end{tabular}

\label{FixPar}
\end{table}

\subsection{The dark matter density profile}\label{DMDEN}

Cosmological simulations suggest that the dark matter halo is well described by a NFW profile \citep{1996ApJ...462..563N}. In our mass model, we choose a generalisation of this density profile based on an axisymmetric two power-law with scale density $\rho_{0}^{\rm h}$, inner slope $\alpha$, outer slope $\beta$, and scale length $a_1$:
\begin{equation}
\rho_{\rm h}\var{s} = \rho_{0}^{\rm h} \left( \frac{s}{a_1} \right)^{-\alpha} \! \left(1 + \frac{s}{a_1} \right)^{\alpha-\beta} ~,
\end{equation}
constant over ellipsoids of equation \ref{s} with axis ratio $q_{\rho} = q_{\rho}^{\rm h}$. When $ q_{\rho}^{\rm h} = 1$ the halo
has spherical symmetry and $s$ is equal to the Galactocentric Spherical
radius $r$. This model is reduced to a NFW when $\alpha=1$ and $\beta=3$. 

In our model, we keep $\alpha$ as a free parameter, and we do not assume any knowledge of its distribution by choosing a uniform prior in the range $\alpha \in\Range{-3}{3}$. This prior gives sufficient freedom to fit the observations without significantly restricting the posterior distribution. The scale length $a_1$ characterises the transition between the inner and the outer slope of the dark matter density profile. We take this scale length as a free parameter following a uniform prior in the range $a_1\in\Range{0}{100}$ kpc. The outer slope $\beta$ defines the shape of the dark matter halo for $R \gg a_1$. Observations of the Milky Way's circular velocity narrow its possible range of values. They exclude $\beta < 2$ to avoid raising rotational curves, as well as $\beta \GtrSim 6$ to avoid rotational curves decreasing too fast. We limit $\beta \in \Range{0}{6}$ using a uniform prior to avoid extreme values of the distribution for computational reasons (see Section \ref{SM}). Even so, $\beta$ is almost unconstrained within this range because our main constraints of the halo, the rotational curve (see Section \ref{RC}) and the stellar streams (see Section \ref{STR}), only introduce constraints for $R\LessSim a_1$. Assuming that $\beta$ is strongly correlated with the mass of the halo, it can be constrained by measurements of the total mass of the Galaxy.

In the cosmological context, dark matter halos are characterized by the
virial mass $M_{\rm vir}$, defined as the mass inside a radius
$r_{\rm vir}$ within which the mean density is $\Delta_{\rm c}$ times
larger than the critical density of the universe:
\begin{equation}
\rho_{\rm crit} \equiv \frac{3H_0^2}{8\pi G} = 140 \,\, {\rm M}_{\Sun} 
\, {\rm kpc}^{-3} ~,
\end{equation}
where we use a Hubble constant $H_0=71$ km s$^{-1}$ Mpc$^{-1}$. For $\Delta_{\rm c} = 200$, we set the dark halo virial mass as $M_{200}^{\rm h}\equiv M_{\rm vir}$, and $r_{200}$ as the radius that solves the equation:
\begin{equation}
M_{200}^{\rm h} \equiv \frac{4\pi}{3} \:\! r^3_{200} \:\! \Delta_{\rm c} \:\! \rho_{\rm crit} = 4\pi q_{\rho}^{\rm h} \! \int_0^{r_{200}} \! s^2\rho_{\rm h} \var{s} \:\! \diff s ~.
\end{equation}

Several methods have been applied to infer the Milky Way mass using the properties of luminous populations, such as the Milky Way's satellites or the kinematics of various dynamical tracers of the Galactic halo \citep[see][ for a review article]{2020SCPMA..6309801W}. In general, these studies use observational data contained in the inner region of the Galaxy. In order to compute the virial mass, they require extrapolations to the virial radius which is about $r_{200}\Approx200$ kpc for the Milky Way. Instead, \citet{2019MNRAS.484.5453C} use the phase-space distribution of the classical satellites of the Milky Way, which are spanned from 50 to 250 kpc from the Galactic centre, to estimate the total mass of the Galaxy:
\begin{equation}
M_{200} \equiv M_{\Star} + M_{200}^{\rm h}~,
\end{equation}
where $M_{\Star}$ is the total baryonic mass. Our model includes the mass of the bulge, thin, and thick disc, thus $M_{\Star} = M_{\rm b} + M_{\rm d}^{\rm n} + M_{\rm d}^{\rm k}$. In order to constrain the slope $\beta$, we include in the likelihood function the measurement of \citet{2019MNRAS.484.5453C} of the mass within a radius of $r_{200} = 215.3 \pm 12.9$ kpc with symmetrized uncertainties:
\begin{equation}\label{M200mw}
M_{200} = (1.17 \pm 0.21)\pd{12} \,\, {\rm M}_{\Sun} ~.
\end{equation}

By imposing that the density of the dark matter halo is constant over ellipsoids of equation \ref{s}, we have assumed an axisymmetric halo with axis of symmetry perpendicular to the disc. In principle, the Large Magellanic Cloud (LMC) should be the main cause of deviations from an overall axisymmetric Galactic potential (see Section \ref{comMC}). We neglect the LMC in our work. Despite its large mass, recent work suggests that the LMC halo mass may be about 1/5 of the Milky Way halo mass \citep{2020MNRAS.495.2554E, 2021MNRAS.501.2279V, 2021ApJ...923..149S}, the tidal acceleration of the LMC on the stellar streams we analyze is only 1 to 2 per cent of the total Milky Way acceleration, as discussed in Section \ref{comMC}. This is due to the distance of the LMC at $\Approx40$ kpc from the streams. We also assume that any deviations from the axisymmetric configuration of the dark matter halo caused by the tides from the LMC are similarly negligible.

The flattening of the Milky Way's halo has been investigated using different kind of methods. For example, constructing self-consistent models of the Galaxy assuming that the distribution of stars in the halo or the globular clusters are in equilibrium. Stellar streams has also been used for this purpose, specially the Sagittarius stream, GD-1, and Palomar 5. We provide a detailed compilation of all these measurements in Section \ref{comOBS}. On the other hand, cosmological simulations statistically predict the shape of the dark halos of Milky Way-like galaxies. In general, simulations that only use dark matter obtain prolate triaxial halos. The introduction of baryons and several feedback effects produce significantly rounder halos. A detailed exposition of these results and a comparison with observational measurements is included in Section \ref{comCS}. Here, we take the axis ratio as a free parameter following a uniform prior large enough not to significantly restrict the posterior distribution in the range $q_{\rho}^{\rm h} \in \Range{0}{6}$.

We also take the scale density $\rho_{0}^{\rm h}$ as a free parameter because it cannot be directly constrained. We adopt a uniform prior in the range $\rho_{0}^{\rm h} \in \Range{0}{1.5}\pd{8}$ ${\rm M}_{\Sun}\,{\rm kpc}^{-3}$ to cut larger values for computational purposes (see Section \ref{SM}). All the parameters of the halo are specified in Table \ref{FreePar}.

\section{Kinematical and dynamical constraints}\label{KCons}

 In addition to the priors derived from observed star distributions and
mass estimates introduced in Section \ref{MassModel}, we include more
detailed kinematical and dynamical constraints from observations in the
solar neighbourhood and the local disc: the position and velocity of the
Sun, the proper motion of Sgr A*, the vertical gravitational
acceleration in the disc at the solar position, and the circular
velocity curve of the Milky Way. These constraints, discussed in
Subsections \ref{KSun}, \ref{Vacc}, and \ref{RC}, are important to reduce the multiple parameter degeneracies of our model potential. We also present in
Subsection \ref{STR} the way we incorporate the additional independent
constraints from the the observations of the stellar streams of
NGC 3201, M68, and Palomar 5.

\subsection{Position and velocity of the Sun}\label{KSun}

 The position and velocity of the Sun are required to determine the
relation between the Galactocentric and Heliocentric coordinate systems.
The distance from the Sun to the Galactic centre is measured to
0.3 per cent accuracy by comparing radial velocities and proper motions of
stars orbiting the Galaxy central black hole Sgr A$^*$
\citep{2019AandA...625L..10G}, $R_{\Sun} = 8.178\pm0.026\, {\rm kpc}$
(including both statistical and systematic error). For the Sun vertical
position, we adopt the central value of the estimate
$z_{\Sun} = 25\pm5$ pc from \citet{2008ApJ...673..864J} (the measurement
error is negligible for our purpose in this case).

 For the Solar velocity with respect to the Local Standard of Rest,
we use the value obtained from the stellar kinematics of the Solar
neighbourhood by \citet{2010MNRAS.403.1829S},
\begin{equation}
v_{\Sun} \equiv
\left(\!\!
\begin{array}{c}
U_{\Sun}\\
V_{\Sun}\\
W_{\Sun}\\
\end{array}
\!\!\right)
=
\left(\!\!
\begin{array}{c}
11.1\pm1.25\\
12.24\pm2.05\\
7.25\pm0.62\\
\end{array}
\!\!\right)
\begin{array}{c}
\\
{\rm km \, s}^{-1} ~,\\
\\
\end{array}
\end{equation}
where $U$ points to the Galactic centre, $V$ is positive along the
direction of the Sun's rotation (clockwise when viewed from the North
Galactic Pole), and $W$ is positive toward the North Galactic
Pole. We take $R_{\Sun}$ and $v_{\Sun}$ as free parameters of our model
with Gaussian priors given by these observational errors, with values
listed in Table \ref{FreePar}, to properly take into account the implied
uncertainties.

 The gradient of the total gravitational potential at the solar position
determines the circular velocity of the Local Standard of Rest (LSR),
$\Theta_0$. The total tangential velocity of the Sun is constrained
by the observed proper motion of the Sgr A$^*$ source, the nuclear black
hole of the Milky Way, measured by \citet{2004ApJ...616..872R}. The
component along Galactic longitude of this proper motion, $\mu_l$, is:
\begin{equation}\label{SgrA}
\mu_l \equiv -{\Theta_0 + V_{\Sun}\over R_{\Sun} } =
 -6.379 \pm 0.026 \,\, {\rm mas \, yr}^{-1} ~.
\end{equation}
We assume that the black hole is located at the Galactic centre and is static, and include this proper motion and error in the likelihood function to constrain our model. The component along Galactic latitude measured in \citet{2004ApJ...616..872R} is consistent with the vertical component of the solar motion $W_{\Sun}$, and the measurement of $R_{\Sun}$, but with a larger error, so we neglect it in our analysis.

\subsection{Vertical gravitational acceleration}\label{Vacc}

 The vertical gravitational acceleration $K_z$ near the disc is used to
constrain the disc surface density, and several studies have obtained
values $K_z \Approx 2 \, {\rm (km/s)}^2 \, {\rm pc}^{-1}$ at
$z \Approx 1$ kpc \citep[e.g.][]{1991ApJ...367L...9K, 2004MNRAS.352..440H, 2013ApJ...772..108Z, 2014AandA...571A..92B}.
\citet{2013ApJ...779..115B} were able to obtain measurements at several
radial distances along the Galactic plane. We do not include these observations because they were obtained assuming a
spherical dark matter halo, and this might introduce an unwanted bias in
our model fit. We use only the measurement by
\citet{2004MNRAS.352..440H} at $z=1.1$ kpc in the solar neighborhood:
\begin{equation}\label{Kz}
\begin{split}
K_{z} &= 2.00\pm0.16 \,\, {\rm km}^2 \, {\rm pc}^{-1} \, {\rm s}^{-2} \\
&= 2\pi G\, (74\pm6) \,\, {\rm M}_{\Sun} \, {\rm pc}^{-2} ~,
\end{split}
\end{equation}
because the large uncertainty will prevent the introduction of a significant bias. We evaluate $K_{z}$ from the potential of our model including the disc and dark halo at $z=1.1$ kpc, and discuss the introduced constraint in Section \ref{sep}.

\subsection{The Milky Way rotation curve}\label{RC}

 The Milky Way rotation curve for $R<R_{\Sun}$ has been measured using
the tangent-point method \citep[see e.g.][]{2006ApJ...641..938L,2007ApJ...671..427M,2016ApJ...831..124M}, and for $R>R_{\Sun}$ using velocities and distances of various tracers \citep[e.g.][]{2012ApJ...761...98K,2014AandA...563A.128L,2016MNRAS.463.2623H}.  These measurements have recently been improved by \citet{2019ApJ...871..120E} with a large sample of red giant stars with 6-dimensional phase-space coordinates obtained by combining spectral data from APOGEE with photometric information from WISE, 2MASS, and \textit{Gaia}. They determine the circular velocity from 5 to 25 kpc with an accuracy characterised by a standard error $\LessSim 3$ km s$^{-1}$ and a systematic uncertainty at the $\Approx 2\,\text{-}\,5$ per cent level of the measurement. Their modeling is compatible with ours to avoid any systematic bias (they assume an axisymmetric potential and approximately the same values of $R_{\Sun}$ and $\Theta_0$ that we use).

We constrain our model using the 38 measurements of the rotation curve of \citet{2019ApJ...871..120E} at different radii. We assume the measurements follow a Gaussian distribution, with a dispersion equal to the symmetrized statistical errors given in \citet{2019ApJ...871..120E}. We add a constant systematic error of 3 per cent, a good approximation in the range $R\Approx\Range{5}{15}$ kpc. The rotation curve with our assumed errors is shown in Section \ref{MRC}.

\subsection{Stellar Streams}\label{STR}

Several tidal streams have been discovered in the Milky Way \citep{2016ASSL..420...87G, 2023MNRAS.520.5225M}, and each of them may provide us interesting constraints on the Galactic potential. The most massive streams in the Milky Way are associated with the Large Magellanic Cloud and the Sagittarius dwarf galaxy. They have been used to study the potential of the Galaxy by numerous authors (see Section \ref{comOBS}). Even though, streams that are thinner and dynamically cold are easier to model to constrain the potential because the stream itself is already a good approximation to a Galactic orbit, and self-gravity and hydrodynamic effects on gas clouds that result in star formation complicate the picture in the massive streams. Some of the thin streams, such as GD-1 and Orphan streams, do not have an identified progenitor. GD-1 is believed to be the remnants of totally destroyed globular cluster \citep{2020MNRAS.494.5315D} and Orphan's progenitor is likely to be a dwarf spheroidal galaxy \citep{2023ApJ...948..123H}. The lack of a progenitor makes these streams more difficult to model but they can also be useful as the data improve. When a progenitor with a measured distance and kinematics is known, its orbit eliminates degeneracies to create a phase-space model of the stream.

In this work, we will focus on streams generated by globular clusters, and we will use only three of them: the streams of the globular clusters NGC 3201 \citepalias{2021MNRAS.504.2727P}, M68 \citepalias{2019MNRAS.488.1535P}, and Palomar 5 \citep{2001ApJ...548L.165O}. These streams are chosen because their progenitors have a phase-space position measured with high precision, and they are long, thin and dynamically cold. These characteristics make it easier to determine the orbit of the progenitor than other streams generated by globular clusters with a more complex morphology, such as the Omega Centauri stream \citep{2019NatAs...3..667I}. Moreover, it has been possible to discover a particularly large number of member stars in the \textit{Gaia} catalogue compared to other globular cluster streams that are generally more distant from the Sun, such as Palomar 13 \citep{2020AJ....160..244S} or NGC 5466 \citep{2021MNRAS.507.1923J, 2022MNRAS.513..853Y}. In the case of Palomar 5, there are also several radial velocities that add useful information. In our previous papers \citepalias{2019MNRAS.488.1535P,2021MNRAS.504.2727P}, we showed how reliable stream members can be identified in the \textit{Gaia} catalogue and used to obtain a model of the streams for M68 and NGC 3201. In the latter case, we substantially extended the known length of the stream and demonstrated the importance of correcting for dust absorption to check for consistency of the photometry with the globular cluster H-R diagram. In this paper, we will also obtain a list of highly likely members of the Palomar 5 stream obtained from the \textit{Gaia} catalogue. These combined 3 streams will then be used to fit a best model for the Galactic potential, together with all other constraints discussed above.

There are other globular clusters with a known stream with similar characteristics to the cases studied. The best example is the M5 (NGC 5904) stellar stream. A section of the trailing arm of this stream has been observed extending along 50 deg in the sky, and about 70 stars from the \textit{Gaia} catalogue have been identified as likely members \citep{2019ApJ...884..174G}. Furthermore, most globular clusters should have associated stellar streams, so many more will be discovered in the future. As many streams as possible should be added to improve the analysis and modelling we do in this paper.

\subsubsection{NGC 3201 stellar stream}

 The stellar stream of NGC 3201 was initially discovered by
\citet{2019ApJ...872..152I} and was named Gj\"oll, without identifying it
with its progenitor NGC 3201. The identified stream was actually a
section of the trailing arm, moving behind the cluster. The extent of
the stream was revealed to be much larger, and was identified with the
tidal stream of the globular cluster NGC 3201 by \citetalias{2021MNRAS.504.2727P}. Part of the stream is not easily
observable because it is projected behind the Galactic disc, strongly obscured by dust and with a high density of foreground stars. This makes the selection of member stars difficult, mostly in the leading arm and near the globular cluster.

In this paper, we use a subset of 54 \textit{Gaia} stars that were found to be highly likely stream members in the study of \citetalias{2021MNRAS.504.2727P}. This subset limits the stream to the region defined by the right ascension $65<\alpha<120$ deg. This region excludes the areas deeply obscured by dust and with the highest foreground contamination. It also excludes the stars located in the outermost part of the cluster, where the separation between bound and escaped stars cannot be precisely established. In this region, a section of the stream can be identified by applying cuts in phase-space coordinates and selecting the stars compatible with the H-R diagram of NGC 3201. The separation of the stream stars from the foreground is very effective in this region because the stream is close to the Sun, and its proper motions are significantly larger than those of the star foreground. Along this section of the stream, the stream membership can be asserted on a star-by-star basis without relying on a statistical determination using the potential of the Galaxy and a density model of the stream. In this way, by limiting the extension of the stream to the section where we can see it clearly, we eliminate possible biases introduced by the selection method towards a particular Galactic potential, as we might introduce if we use the entire sample of stars in \citetalias{2021MNRAS.504.2727P}. We include in Appendix \ref{AppA} a detailed description of the section of the stream we use to constrain the Galactic potential, and of the star selection process.

\begin{figure*}
\includegraphics[]{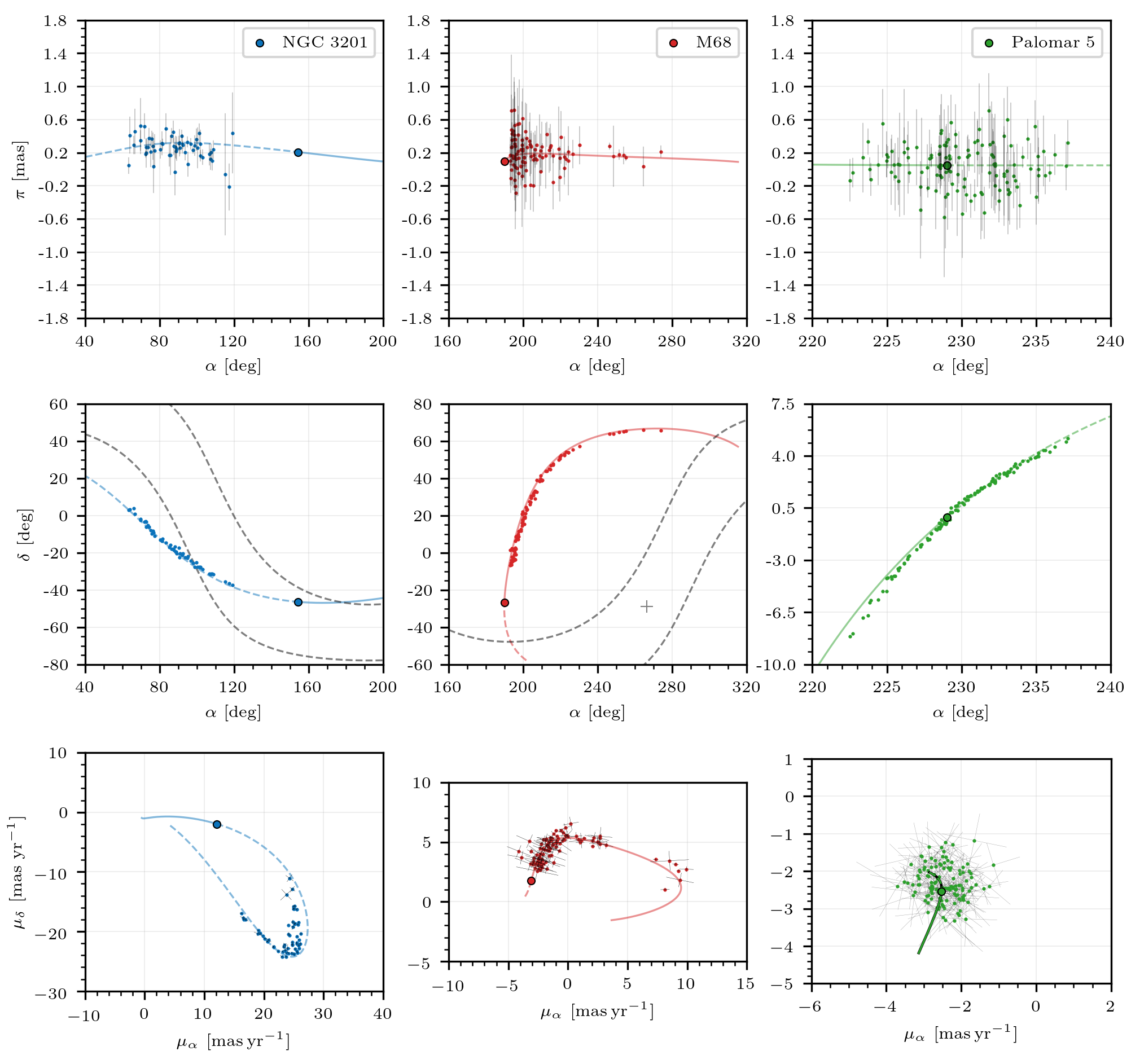}
\caption{Phase-space position of stream stars from the GDR2 catalogue (points with error bars) for globular clusters NGC 3201 (blue),
M68 (red), and Palomar 5 (green). Big dots show the current phase-space
position of clusters and lines show their orbits forwards (solid) and
backwards (dashed) during 60 Myr computed using the best-fitting orbit.
\textit{Middle panels:} The grey dashed lines mark the Milky Way disc
limits at $b=\pm 15$ deg and the grey cross the Galactic centre.}
\label{streams_ICRS}
\end{figure*}

Figure \ref{streams_ICRS} shows the parallax $\pi$, declination $\delta$, right ascension $\alpha$, and proper motion components $\mu_\delta$ and $\mu_{\alpha *}\equiv \mu_\alpha\cos\var{\delta}$ of the stream stars we use to constrain the potential of the Milky Way. The small dots represent the 54 stars, and the large dot marks the position of the globular cluster NGC 3201. The black dashed lines in the $\alpha-\delta$
diagram indicate the region within 15 degrees of the Galactic plane, and
the colored curve is the best-fitting orbit of the globular
cluster, showing an integration time of $60\, {\rm Myr}$ backward
(dashed line) and forward (solid line) in time. The stream spans about 60 degrees on the southern Galactic hemisphere and is located close to the Galactic disc, and comes to a closest distance of 3 to 4 kpc from the present position of the Sun. The stream stars that are passing close to us have relatively large proper motions of
$\Approx 20$ ${\rm mas\,yr}^{-1}$, which facilitate their identification
and makes them useful for kinematic studies using \textit{Gaia} proper motions.
Note that the parallaxes are too small to provide much information,
and the useful kinematic information of the streams are the \textit{Gaia} proper
motions.

The kinematics of NGC 3201 are specified in Table \ref{FreePar} and
Table \ref{FixPar}. We use the coordinates of \citet{1996AJ....112.1487H, 2010arXiv1012.3224H}, with negligible errors, and the heliocentric distance $r_{\rm h}$ from the same catalogue assuming a 2.3 per cent uncertainty. The radial velocity $v_r$ is from \citet{2019MNRAS.482.5138B}, who compile several measurements. We use proper motions from \citet{2019MNRAS.484.2832V}, based on GDR2 data. We take these properties as free parameters and take the quoted errors from the observations, listed in Table \ref{FreePar}, as a prior assuming they are Gaussian.

\begin{figure}
\includegraphics[width=1.0\columnwidth]{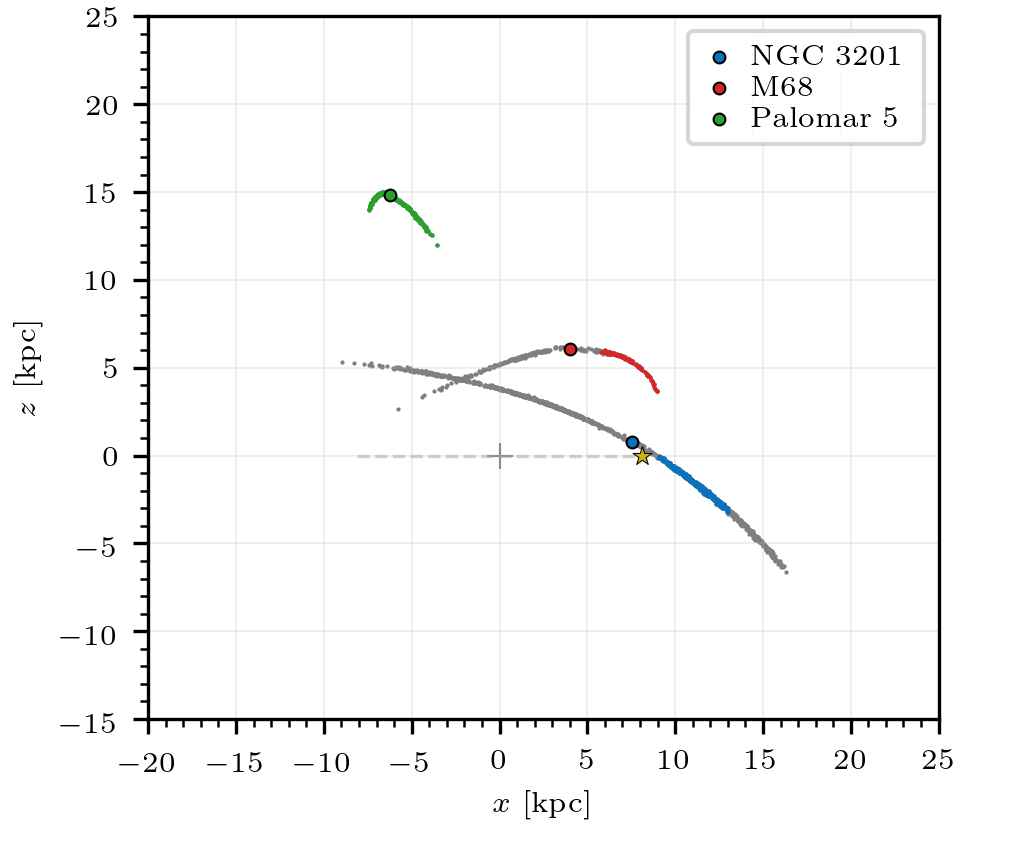}
\includegraphics[width=1.0\columnwidth]{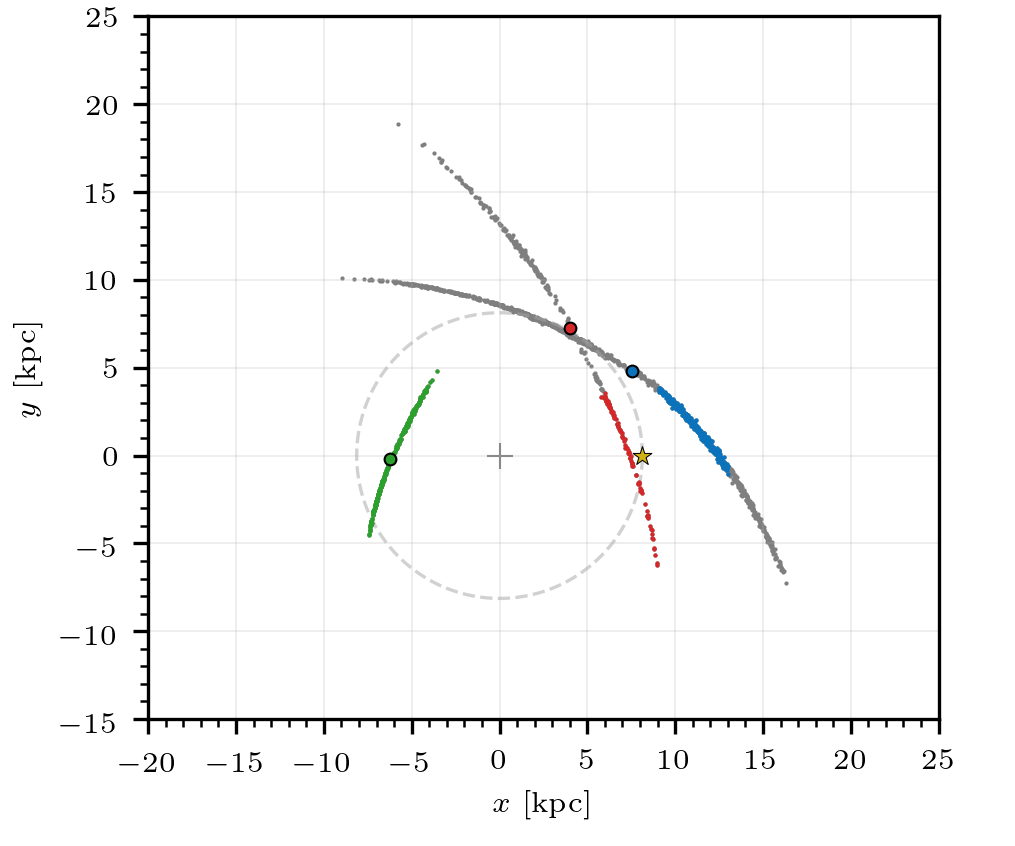}
\caption{Simulated stream stars in Galactocentric coordinates, taking the
stars stripped from the globular cluster during the last 1.5 Gyr for NGC 3201, and M68 and 4 Gyr for Palomar 5. The big dots show the current position of the globular clusters NGC 3201 (blue), M68 (red), and Palomar 5 (green). The coloured stars display approximately the section of the stream that fits with our \textit{Gaia} selection plotted in Figure \ref{streams_ICRS}. The yellow star marks the Sun's position and the grey dashed line shows its orbit assuming a circular motion. The grey cross marks the Galactic centre.}
\label{streams_xyz}
\end{figure}

We use the mean values of the phase-space coordinates of the cluster and a fiducial Galactic potential to simulate this stream (see Section \ref{SM}). We assume that the mass and size of the cluster are fixed throughout the orbit. These properties are listed in Table \ref{FixPar}. In Figure \ref{streams_xyz} we plot in Galactocentric Cartesian coordinates the simulated stars stripped from the cluster during the last 1.5 Gyr, and we highlight in blue the simulated stars that approximately fit with our selection of \textit{Gaia} stars. We also indicate the position of NGC 3201 with a big blue dot. We see that the observed portion of the stream is
located approximately from 10 to 13 kpc from the Galactic centre and very close to the Galactic disc, in the range -3 to 0 kpc.

\subsubsection{M68 stellar stream}\label{str_m68}

The stellar stream associated with the globular cluster M68 (NGC 4590) is a long and thin structure that spans about 190 deg over the north Galactic hemisphere. This stream appears in \citet{2019ApJ...872..152I} named as Fj\"orm without being associated with M68. We use a 98-star subset of the stream candidates selected in \citetalias{2019MNRAS.488.1535P}, corresponding to the stars with $\delta>-8$ deg. With this cut, we exclude stars located close to the Galactic disc, where the correct determination of stream members is uncertain due to the high level of foreground contamination. This selection includes stars along almost the entire leading arm of the stream which appears projected onto the halo. This section is described in detail in the Appendix \ref{AppA}, and the final selection of stars we use to constrain the Galactic potential is shown in red in Figure \ref{streams_ICRS}. Most of these stars are located very close to the Sun at $\Approx 5.5$ kpc and have proper motions approximately in the range $\Range{5}{10}$ ${\rm mas\,yr}^{-1}$ making them easily identifiable with respect to the foreground. On the other hand, the section closer to the globular cluster and all the trailing  arm are completely obscured by foreground stars, most of them belonging to the disc. Similarly to the stream of NGC 3201, we can assert the stream membership of each star by direct inspection of the stars passing a set of cuts in phase-space, colour and magnitude. This avoids possible biases towards the potential used by the statistical method in \citetalias{2019MNRAS.488.1535P}.

For M68, we also take its sky coordinates as fixed parameters and the remaining phase-space coordinates as free parameters, assuming a 5 per cent of uncertainty for the heliocentric distance. We list their values in Table \ref{FixPar} and in Table \ref{FreePar} respectively. In Figure \ref{streams_xyz} we observe that the stream is located at about 9 to 12 kpc from the Galactic centre and about 4 to 6 kpc from the Galactic disc.


\subsubsection{Palomar 5 stellar stream}

The Palomar 5 tidal tails were discovered by \citet{2001ApJ...548L.165O} by noticing an excess of stars around the globular cluster using photometric data provided by Sloan Digital Sky Survey. Further work improved the definition of the tidal tails and extended its length up to 23 deg in the sky \citep[e.g.][]{2012ApJ...760...75C}. Its full phase-space distribution has been described by the identification of individual stars in the tidal stream \citep[e.g.][]{2015MNRAS.446.3297K, 2016ApJ...819....1I, 2017ApJ...842..120I}, and improved using the GDR2 catalogue \citep{2020MNRAS.493.4978S, 2019AJ....158..223P}.

\begin{figure}
\includegraphics[width=1.0\columnwidth]{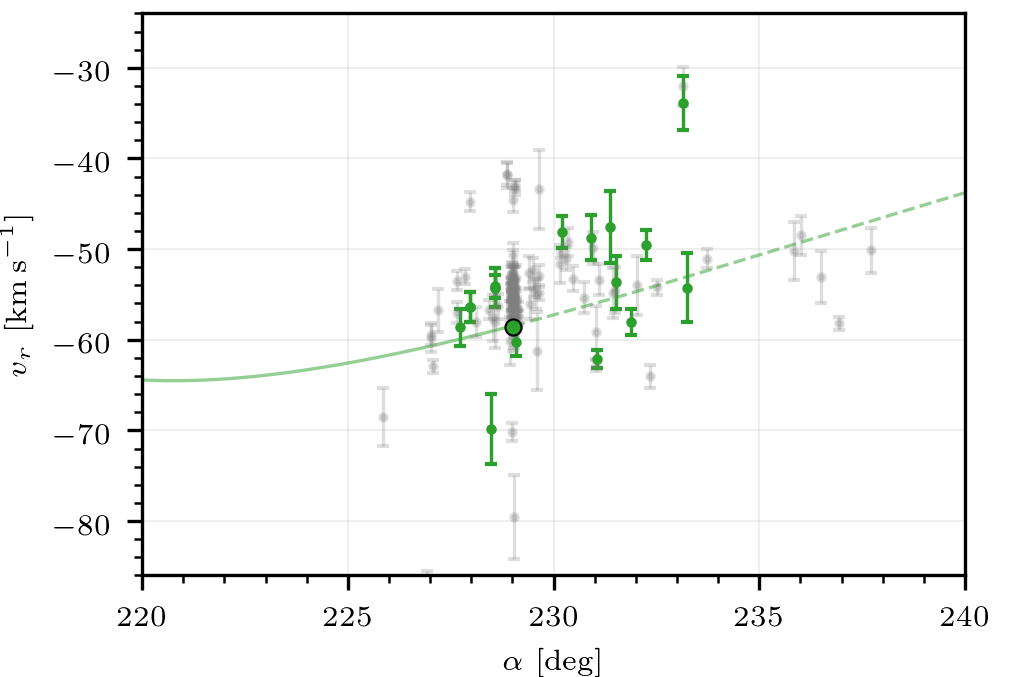}
\caption{Right ascension and radial velocity of the stream stars of Palomar 5 from \citet{2017ApJ...842..120I} (grey) with error bars. The green points mark the 15 stars that match with our \textit{Gaia} selection. The big dot shows the current position of the cluster and the lines show its orbit forwards (solid) and backwards (dashed) during 60 Myr computed using the best-fitting values of the free parameters.}
\label{rv_pal5}
\end{figure}

In this paper, we use our selection of stream stars made following the method described in \citetalias{2019MNRAS.488.1535P}. This method is based on a maximum-likelihood technique designed to distinguish stars compatible with being tidally stripped from a known globular cluster. These stream stars appear as an overdensity that is statistically identified when compared to a phase-space model of the Milky Way. The stars that most likely belong to the stream are selected by choosing those with the largest intersection with a phase-space density model of the stream. This model is computed numerically by optimising several free parameters to maximise the intersection between the stellar overdensity and the model. The free parameters include the potential model of the Galaxy and the heliocentric distance and velocity of the cluster within the constraints of the available observations. The stars selected using phase-space information are consistent with the recent observations of the Palomar 5 stream by \citet{2020ApJ...889...70B} using $grz$ photometry from DECaLS, which includes stars up to 24 mag. This ensures that our selection methodology does not introduce any bias that could affect the determination of the Galactic potential. Our final selection only includes the stars that are colour and magnitude compatible with the H-R diagram of the progenitor cluster. We show in green the 126 selected stars in Figure \ref{streams_ICRS}. We list the phase-space coordinates, colours and magnitudes and explain the details of the selection procedure in Appendix \ref{E}. None of the selected stars has radial velocity in the GDR2 catalogue, but 15 of them match with stars with radial velocity measured by \citet{2017ApJ...842..120I}. We take their measurements, list them in Appendix \ref{E}, and display the radial velocity in function of right ascension in Figure \ref{rv_pal5}.

The Palomar 5 tidal stream is projected onto the halo just over the Galactic centre. Our selection covers about 16 deg in the sky, almost the entire stream. We observe a well-defined structure, with two long and thin arms connected to the globular cluster. The observations of \citet{2020ApJ...889...70B}, show a low surface-brightness extension of $\Approx10$ deg on the trailing arm. Our selection does not include this extension, as the \textit{Gaia} $\Gband$ magnitude limitation of $G<21$ mag makes it difficult to identify stars in the trailing arm faint extension. In the proper motion space, we observe a bunch of stars. We do not observe the long and thin shape of the stream because the internal dispersion of proper motions of the stream stars is much smaller than the \textit{Gaia} observational uncertainties.

For the phase-space coordinates of Palomar 5, we take the values from the same references as in the previous cases (see Table \ref{FreePar} and \ref{FixPar}), except for the heliocentric distance $r_{\rm h}$ taken from \citet{2019AJ....158..223P}. In general, the measurements of $r_{\rm h}$ approximately range from 20 to 23 kpc, here we use $20.6\pm0.2$ kpc. The simulation of the stream (green dots in Figure \ref{streams_xyz}) shows that the stream is located at about 13 to 17 kpc from the Galactic centre and about 12 to 15 kpc from the Galactic disc.

\section{Statistical methodology}\label{SM}

 Given a set of free parameters $\theta$ and a set of observational
measurements $d$, the posterior distribution of all parameters together $p\var{\theta|d}$ can be determined by the Bayes' theorem:
\begin{equation}\label{posF}
p\var{\theta|d} = \frac{\mathcal{L}\var{d|\theta} \, p\var{\theta}}{p\var{d}} ~,
\end{equation}
where $\mathcal{L}\var{d|\theta}$ is the likelihood function,
$p\var{\theta}$ is the prior distribution of all the parameters, and
$p\var{d}$ is a normalisation constant. In our model, we
use 4 free parameters that characterise the position of the Sun, 12 for
the potential of the Milky Way, and 4 for the phase-space position of
each globular cluster. The free parameters, including their prior
distribution functions, are described in Section \ref{MassModel} and
listed in Table \ref{FreePar}; those without a specified prior in this
table are assigned a flat prior (with fixed limits added for numerical
convenience that are broad enough to have no impact on our results). Parameters that are kept fixed are listed in Table \ref{FixPar}.

The likelihood function is computed as the product of the likelihoods associated to each observational constraint. This is divided into two sets of data: first, the traditional dynamical constraints from equilibrium models of the Milky Way, consisting of a total of 5 measured variables described in Sections \ref{MassModel}, \ref{KSun}, and \ref{Vacc}
$(f_\varSigma, M_{\rm b}, M_{200}, \mu_{l}, {\rm and}\, K_{z})$, and the 38 values of the velocity rotation curve of \citet{2019ApJ...871..120E}, described in Section \ref{RC}. For this, we assume a Gaussian distribution for all of these 43 variables and treat them independently, even though the 38 points of the rotation curve have a correlated error. We simply use error bars for the rotation curve that are larger than the purely statistical errors, by adding a systematic error of 3 per cent to each point, which roughly compensates for the error correlations.

The second data set are the observations of the 3 streams used in this paper. The data consist of a list of the phase-space coordinates of
stars from the \textit{Gaia} catalogue that have been selected as stream
members. These include positions (with negligible errors), proper
motions and parallaxes with the covariance matrix of the GDR2 measurement errors. In addition, radial velocities and their errors
are available only for part of the stars of the Palomar 5 stream. We
define the likelihood as the convolution of this measurement error
distribution with a phase-space probability density model of the stellar
stream (see Section \ref{PSMSS}). A more detailed definition of the
likelihood function is explained in Appendix \ref{A}.

To obtain the posterior distribution of the parameters of our model, we use the Metropolis-Hastings algorithm \citep{MacKay2003}, which is a Markov Chain Monte Carlo method that generates random samples following a probability density function. We use our own implementation of this method based on a Gaussian transition distribution with adjusted covariance matrix to maximize performance, and run 72 walkers initialized with a random position. The algorithm converges to a stationary set of samples after about $10^5$ steps for all parameters. These steps have been excluded to avoid a bias due to the random initial configuration, and the posterior distributions are drawn using the next $5\times10^5$ steps of the chain. The halo parameters $\rho_0^{\rm h}$, $a_1$, and $\beta$ present an asymmetric posterior distribution with an extended tail towards large values (see Section \ref{halo}). To ensure convergence, we limit the tail of the distributions with the boundaries of the uniform priors introduced in Section \ref{DMDEN}. We find the best-fitting values using a Nelder-Mead Simplex algorithm, and present the results in Section \ref{Res}.

\subsection{Phase-space model of the stellar stream}\label{PSMSS}

The phase-space probability density model of the stellar stream is constructed from simulated particles escaping from the globular cluster, modeled as a static Plummer potential orbiting the static Milky Way potential, subject to the tidal forces of the Galaxy. The model depends on the potential of the Milky Way, and the globular cluster mass, scale length and orbit.

Several methods have been developed to quickly simulate stellar streams. For example, the streak-line or particle-spray method avoids calculating the orbit of non-escaping stars (with the small time steps required in the cluster core) by releasing particles from the Lagrange points \citep[e.g.,][]{2012MNRAS.420.2700K}. Alternatively, some methods rely on the simple structure of the stream in action-angle coordinates to create prescriptions for its phase-space structure \citep[e.g.,][]{2014ApJ...795...95B,2015MNRAS.452..301F}. None of these methods is fast enough to compute a random sample large enough to adequately describe the posterior function (eq. \ref{posF}) in a reasonable time with our computational resources, for the large variety of model parameters we want to examine.

For this reason, we do not simulate a stellar stream for each evaluation of the likelihood function. We do an accurate simulation only initially for fiducial parameter values, and then, we assume that the position and velocity dispersion of the stream with respect to the orbit of the progenitor do not change for small variations of the potential of the Galaxy and the phase-space location of the cluster. This assumption allows us to obtain an approximation of the stream phase-space structure without the computational cost of a numerical simulation. The initial simulation is carried out using the method described in \citetalias{2019MNRAS.488.1535P}. The procedure we apply can be summarized as the following steps:
\begin{enumerate}
 \item We compute the orbit of the globular cluster backwards in time during 1.5 Gyr for NGC 3201 and M68, and 4 Gyr for Palomar 5, starting from the present mean position and velocity. The time intervals are selected to match the size and length of the observed streams. The orbit is computed using the fiducial potential of the Galaxy defined in \citetalias{2019MNRAS.488.1535P}.
 \item We assume the globular cluster is initially in dynamical equilibrium and we randomly generate member stars using the equilibrium distribution function. In our case, we adopt a Plummer sphere, with the core radius $a_{\rm gc}$ and the cluster mass $M_{\rm gc}$ listed in Table \ref{FixPar} for the three globular clusters treated in this paper.
 \item The orbits of the stars are computed starting from the initial position of the cluster centre computed in step (i) plus the distribution of relative positions and velocities computed in step (ii), up to the present time. The stars are treated as test particles moving in the fixed Galactic potential plus the Plummer model potential of constant mass moving along the previously computed orbit.
\end{enumerate}

To compute the stream model for other parameters, we assume that the relative phase-space position of the stream stars with respect to the cluster orbit that we have computed with the fiducial model do not change for small variations of the orbit. We describe the exact procedure in Appendix \ref{B}, and summarize it with the following steps:
\begin{enumerate}
\item We select a section of the cluster orbit corresponding to the cluster motion during 60 Myr for NGC 3201 and M68, and 40 Myr for Palomar 5, backwards and forwards in time with respect to the present location of the cluster. We define a Frenet-Serret trihedron on each point of the orbit. Each trihedron is a orto-normal vector basis defined by the normalised velocity and acceleration of the cluster and their perpendicular vector. This trihedron defines a reference frame with origin on its corresponding point of the orbit.
\item We assign to each star the trihedron located in the closest point of the orbit to the star, determined with a Euclidean distance. We store the position and velocity of the star with respect to the reference frame defined by its trihedron, and assume that this relative phase-space position does not significantly change for small variations of the cluster orbit. We also store the time position of the origin of the trihedron along the section of the orbit of the cluster.
\item For each new evaluation of the likelihood function for different parameters, we compute the new cluster orbit section over the same time period. In the time positions along the orbit previously stored, we compute their corresponding new trihedrons. Finally, we locate each star at the previously stored relative position and velocity with respect to new trihedrons.
\end{enumerate}
This method neglects variations of the internal structure of the stream with respect to the cluster orbit, and changes the stream only due to the variation of the cluster orbit with the potential. Because the stream sections we study are thin and depart only at levels of few percent from the cluster orbit, we expect the error introduced by the method is negligible for our purpose in this paper (see Appendix \ref{B} for a detailed justification of this assumption).

The positions and velocities of the test stars are finally converted to the heliocentric reference frame $(\pi, \delta, \alpha, v_r, \mu_\delta, \mu_\alpha)$, and compared to the observational data. The phase-space probability density model of the stream is constructed in these coordinates using a kernel density estimation method, with a Gaussian distribution as a kernel. We locate the mean of the kernel distributions in the position of the simulated stream stars, and we compute their covariance matrices from the distribution of neighboring stream stars. We describe this method in detail in Appendix \ref{A}.

\section{Results for each model}\label{Res}

 The stellar streams of NGC 3201 and M68 are located at similar distances from
the Galactic centre, covering a range from $r\Approx6$ to 13 kpc.
However, whereas the M68 stream is observed along an orbit portion that
remains $\Approx 5$ kpc above the disc, the NGC 3201 stream traverses the
disc from North to South. In contrast, the Palomar 5 stream is further
away, $\Approx$ 16 kpc from the centre and $\Approx$ 14 kpc above the disc.
Each stream is therefore probing different regions of the Galactic
gravitational potential. To better understand the constraints provided
by each stream, we first present the three mass models obtained by
fitting each individual stream and then the model including all three
streams together.

 The results of our fits are listed in
Table \ref{table_res1} in Appendix \ref{C}, as the median value and
1$\sigma$ error of the posterior distribution marginalized over all
other parameters, for each separate stream and for all streams together. Models with a single stream have a total of 20 free parameters. The model with all streams together requires 28 parameters because each stream includes four free parameters for the phase-space position of the globular cluster. We also list in the table results for other derived
properties of the model, including rotation curve velocities at different radii, $f_\Sigma$, $\mu_l$, the masses of each Milky Way component, and several properties of the dark halo.

\subsection{Consistency with model priors and other observational Data}
\label{sep}

  Our main goal is to obtain new constraints on the mass distribution of
the Milky Way halo, so we will discuss the results for the
five parameters of our halo model: $\rho_0^{\rm h}$, $\alpha$, $\beta$, $a_1$,
and in particular the axis ratio $q_\rho^{\rm h}$. Before this, we briefly
comment on the consistency of all other parameters of our resulting fits
with the priors that are imposed from observational determinations as
discussed in Sections \ref{MassModel} and \ref{KCons}. As seen in Table \ref{table_res1}, all parameters are generally within the errors of the priors, indicating that our models are fully consistent with all these observational constraints and can adequately fit them together with our new conditions from the stellar stream members. This gives us confidence on the results and errors obtained for the dark halo parameters.

  We comment on some of the input parameters that show moderate
discrepancies from the priors. For the Sgr A$^*$ proper motion, all our best-fitting models differ by less than $1.8\sigma$
from the observed value in equation \ref{SgrA}. The rotational velocity of our models at the Solar radius are
$\Theta_0 + V_{\Sun} \Approx 245$ ${\rm km}\,{\rm s}^{-1}$, and
$\Theta_0 \Approx 231\, {\rm km}\, {\rm s}^{-1}$. Comparing these
to other recent observational data, we see that our values are
consistent with determinations of \citet{2019ApJ...885..131R} from
parallaxes and proper motions of molecular masers associated with young high-mass stars:
$\Theta_0 + V_{\Sun} = 247\pm4$ ${\rm km}\,{\rm s}^{-1}$ and
$\Theta_0=236\pm7$ ${\rm km}\,{\rm s}^{-1}$, for
$R_{\Sun}=8.15\pm0.15$ kpc.
They are also similar to \citet{2019ApJ...870L..10M}, who used
classical Cepheid proper motions and radial velocities from
\textit{Gaia} to infer: $\Theta_0 + V_{\Sun} = 246.9\pm1.6\,
{\rm km}\,{\rm s}^{-1}$ and $\Theta_0=233.6\pm2.8\,
{\rm km}\,{\rm s}^{-1}$ for $R_{\Sun}=8.122\pm0.031$ kpc.

 The parameters describing the bulge and disc of the Milky Way are
mostly constrained by our priors described in Sections \ref{MassModel}
and \ref{KCons}. The mass of the thin disc is an exception because the total
disc mass is the main quantity that is degenerate with halo parameters,
and it needs to be constrained by the combination of
rotation curve data and our stream conditions (the thick disc contains
less mass and is therefore less important for the potential, so it is
mostly constrained by the priors). When the M68 stream is used
individually, a larger disc mass by $\Approx 20$ per cent is required compared to the other two streams, which gives a total baryonic mass of $(8.3\pm0.49)\pd{10}$ M$_{\Sun}$. This value is significantly larger than the estimated in other models, e.g. $(6.43\pm0.63)\pd{10}$ M$_{\Sun}$ in \citet{2011MNRAS.414.2446M} or $(7.25^{+0.39}_{-0.68})\pd{10}$ M$_{\Sun}$ in \citet{2020MNRAS.494.4291C}. The combination of M68 and Palomar 5 allows for reducing the parameter degeneracy of the disc mass with the oblateness and density profile of the halo dark matter. Including all the streams together, the resulting model also prefers a similarly high disc mass. Note that the increased disc mass of the models including the M68 stream, results in a larger vertical acceleration $K_z$, with a $2\sigma$ deviation from the observational prior we are using (eq. \ref{Kz}).

The phase-space locations of NGC 3201 and M68 are consistent with
our priors from observations. The case of Palomar 5, on the other hand,
shows discrepancies in the distance from the Sun $r_{\rm h}$ of $2.4\sigma$,
and in the proper motion components of $1.5\sigma$ and $3.1\sigma$. This
may partly be due to systematic errors in the distance measurement of
$20.6\pm 0.2$ kpc in \citet{2019AJ....158..223P} that we use as a prior. Our estimate of $21.19\pm0.15$ kpc is closer to the literature average of $21.9\pm0.5$ kpc from \citet{2021MNRAS.505.5957B}. The large shift preferred by our stream model fit in the proper motion of Palomar 5 may not be entirely explained by the positive correlation with $r_{\rm h}$, and may indicate an inability to obtain a sufficiently good fit to the stream with the model parameterization we have chosen.

\subsubsection{Circular velocity curve}\label{MRC}

\begin{figure}
\includegraphics[width=1.0\columnwidth]{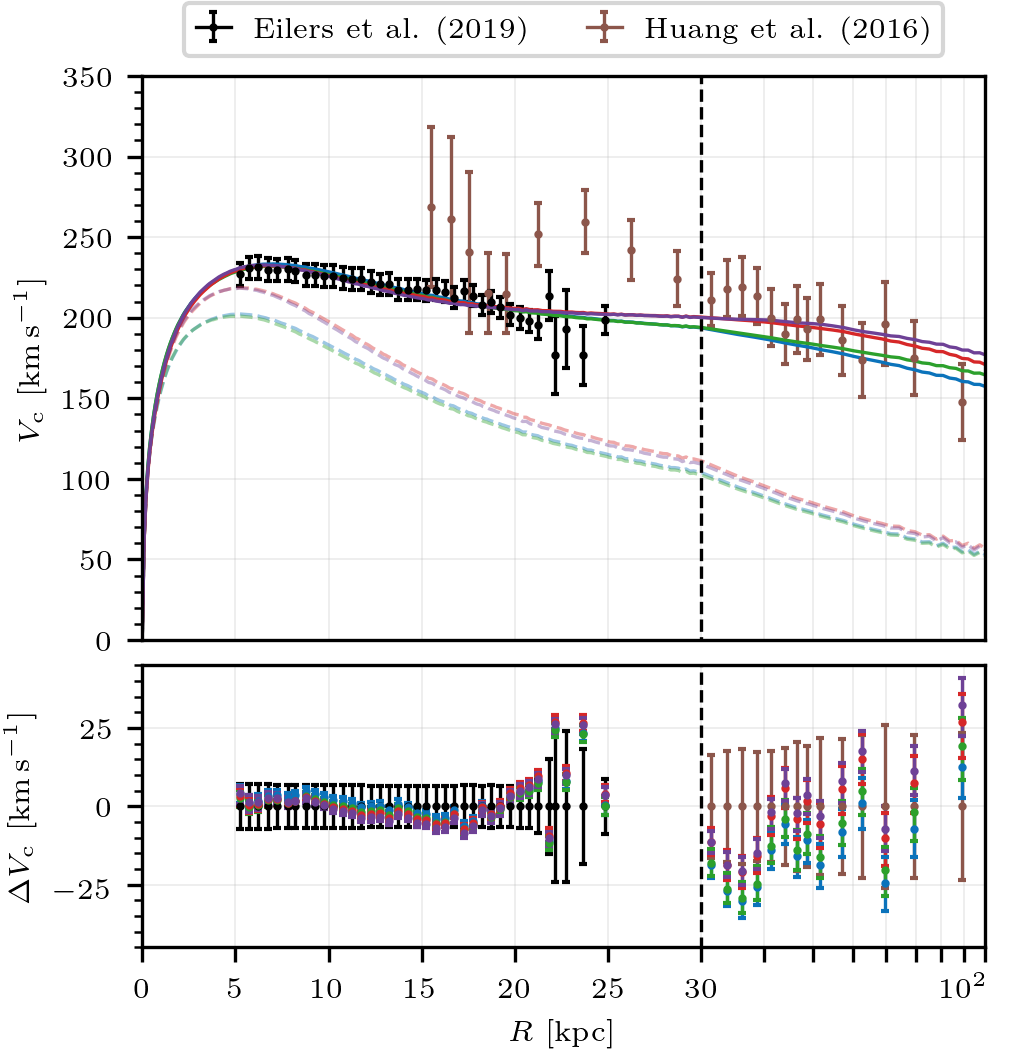}
\caption{Circular velocity curve of the Milky Way. \textit{Top panel:}
Solid lines are best-fitting models rotation curves of the single streams of
NGC 3201 (blue), M68 (red), Palomar 5 (green), and all streams together
(purple). Dashed lines are baryonic component contributions. Black dots
with error bars are rotation curve data by \citet{2019ApJ...871..120E},
with errors computed as described in Subsection \ref{RC}, and brown dots
with error bars are from halo K giant velocities from
\citet{2016MNRAS.463.2623H} (from their HKG sample). Horizontal scale
changes from linear to logarithmic at dashed vertical black line.
\textit{Bottom panel:} Residuals between models and observational data.
}
\label{RCfig}
\end{figure}

 We plot the circular velocity curve of the Milky Way in the top panel
of Figure \ref{RCfig}. Solid lines give the total circular velocity of
the three fitted models of each globular cluster stream, and dashed
lines are the circular velocity of the baryonic mass models only.
Black dots with error bars are the data taken from
\citet{2019ApJ...871..120E} with errors computed as described in
Subsection \ref{RC}. The bottom panel shows the residuals between models
and data. There are no significant differences among the model rotation
curves, which are all consistent with observations.
Since the results of \citet{2019ApJ...871..120E} do not extend to
$R>25$ kpc, we include as brown dots with error bars the independent
data of \citet{2016MNRAS.463.2623H}, who use halo K giant stars
(the HKG sample in the reference) extending from $\Approx16$ to
$100\, {\rm kpc}$, with typical uncertainty $\Approx 20$
${\rm km}\,{\rm s}^{-1}$. These data are also in reasonably good agreement with our models at
$R\GtrSim 30\, {\rm kpc}$, in particular with the magnitude of the slight
decline of the circular velocity at these large radii.

\subsection{Dark matter halo results: individual streams}\label{halo}

\begin{figure*}
\includegraphics[]{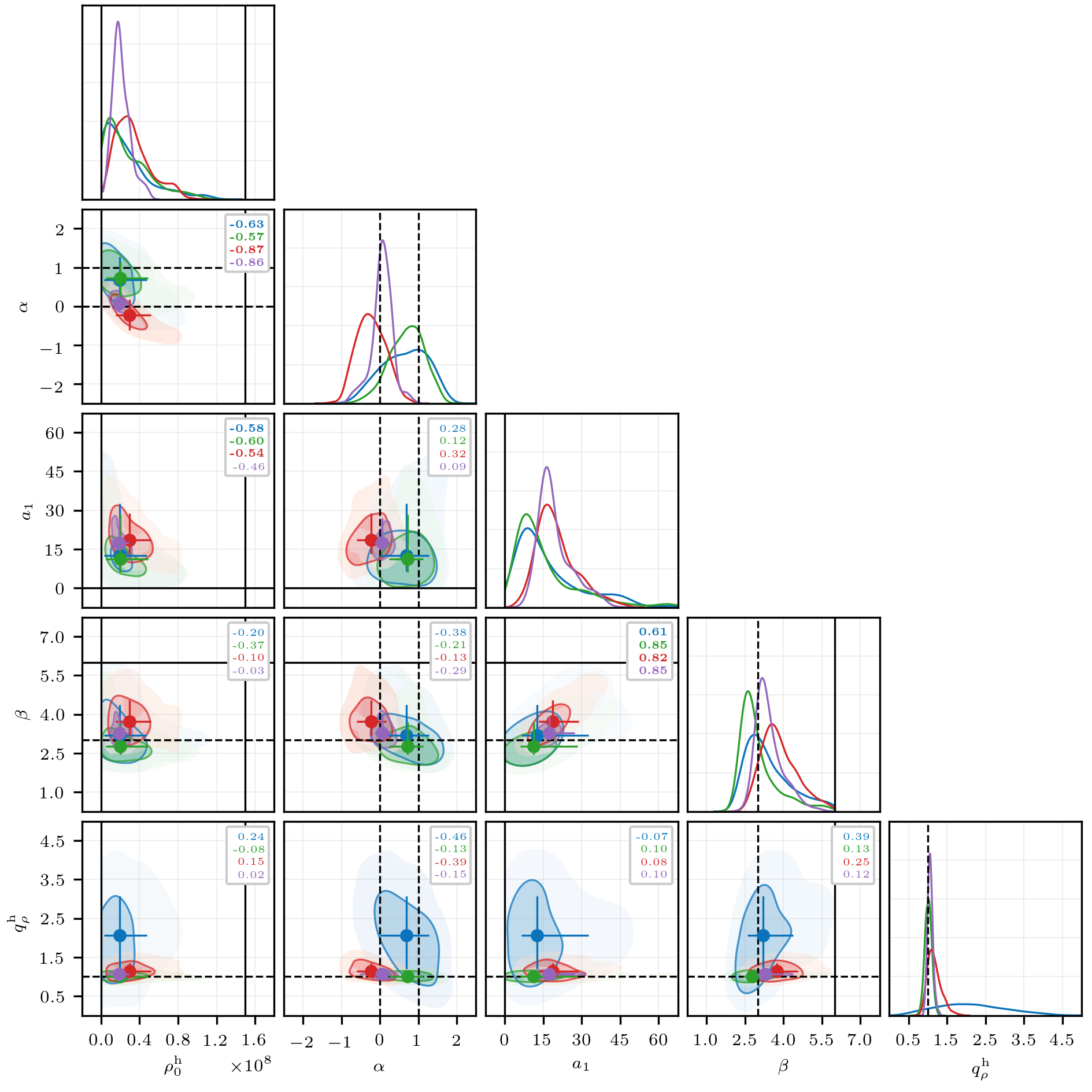}
\caption{Halo parameters corner plot: posterior distributions
marginalised over all parameter pairs (bright shaded areas with solid
contours are $1\sigma$ level, faint shaded areas are $2\sigma$ level),
and each single parameter (curves), for the stream models of NGC
3201 (blue), M68 (red), Palomar 5 (green), and all streams together
(purple). {\it Dots:} distribution medians. {\it Solid black lines:} limits of uniform priors, when present in
displayed intervals. {\it Dashed lines:} Flat and NFW inner and outer
halo slopes ($\alpha=0,1$, $\beta=3$), and spherical configuration
($q_\rho^{\rm h}=1$). Legend of each panel shows Pearson correlation
coefficients.}
\label{corner}
\end{figure*}

 We now present the main results of the paper on the dark matter halo
parameters fitting for each individual stream in this subsection, and all three streams together in the next subsection. In Figure \ref{corner}, we present the posterior distribution function marginalized over each pair of halo parameters (contour panels)
and each individual parameter (colored curves) for each model. We use the same color code as before. We also include the Pearson correlation coefficient between the two marginalised parameters in the legend of each panel.

We first comment on the halo density profile preferred by our models.
In general, all the models demand a flatter density core in the region where baryons dominate ($R \LessSim 15\, {\rm kpc}$) than the NFW density profile. This is reflected in the small values of the inner slope $\alpha$ and in our large core radii $a_1\Approx 15\, {\rm kpc}$ (see Table \ref{table_res1}).
For NGC 3201 and Palomar 5, $\alpha$ is consistent with $\Approx0.7$, and for the M68 stream, the preferred $\alpha$ is actually negative. This result reflects the preference for a more massive disc in this model (implying less dark matter at small radii). A negative $\alpha$ is of course not physical, and is simply indicating the preference for the model for a reduced dark matter density in the central region.

The outer slope $\beta$ is highly correlated with the total mass
$M_{200}$ and $a_1$ in all our models, and it is nearly unconstrained by the stellar streams and rotation curve data. The outer halo density profile is adjusted to fit the constraint of the total mass imposed by
the satellite velocities (eq. \ref{M200mw}). The obtained value is about $\beta=3$ in the three individual stream models.

It is also of interest that the local dark matter density in our
models, $\rho_{\rm h}\var{R_{\Sun}}$, is lower than the value usually
estimated for dark matter detection of $\Approx 0.4$ ${\rm GeV\, cm}^{-3}$
\citep[see, e.g.,][]{2020JPhCS1468a2020D}. The models for NGC 3201 and M68 favor $\rho_{\rm h}\var{R_{\Sun}}\Approx0.2$ GeV cm$^{-3}$, and the more spherical halo of the Palomar 5 model favors a slightly larger value, $\rho_{\rm h}\var{R_{\Sun}}\Approx0.28$ GeV cm$^{-3}$. $\rho_{\rm h}\var{R_{\Sun}}$ depend on several parameters, but mainly correlate with the axis ratio. In general, spherical halos are assumed, which explains the discrepancy with our models requiring prolate halos.

\subsubsection{Dark halo axis ratio}\label{hflat}

\begin{figure}
\includegraphics[width=1.0\columnwidth]{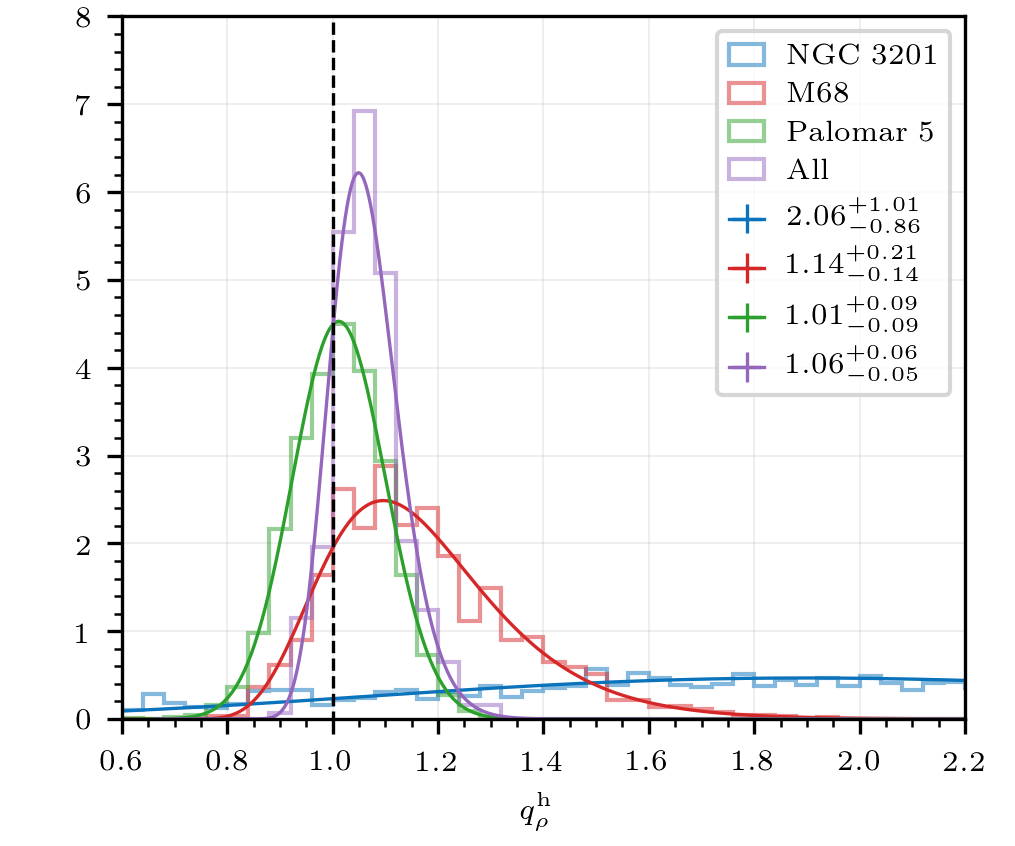}
\caption{
Posterior distribution of the halo axis ratio $q_\rho^{\rm h}$ in the
stream models of NGC 3201 (blue), M68 (red), Palomar 5 (green), and all
streams together (purple), shown as histograms obtained from random
samples. Median and $1\sigma$ levels are shown in the legend, and the
best-fitting log-Normals are shown as solid lines (see Appendix \ref{D}).
Spherical halo ($q_\rho^{\rm h}=1$) is marked as vertical dashed line.
}
\label{q}
\end{figure}

\begin{figure}
\includegraphics[]{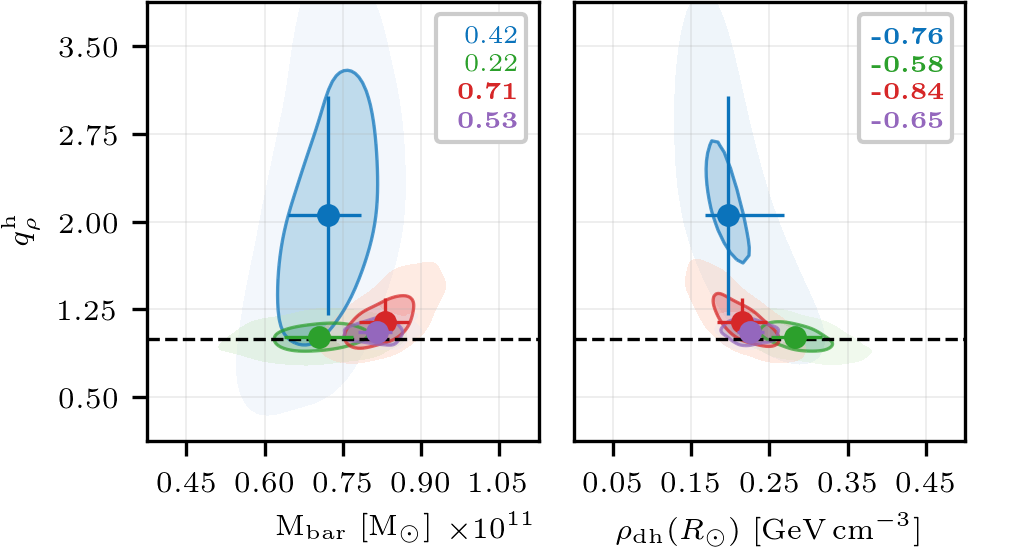}
\caption{Same as Figure \ref{corner}, for the axis ratio distribution
together with the baryonic mass $M_{\Star}$ (left-hand panel) and the
local dark matter density $\rho_{\rm h}(R_{\Sun})$ (right-hand panel).}
\label{corner_Mrho}
\end{figure}

 We now discuss the result on the main focus of our paper, the dark halo
axis ratio $q_\rho^{\rm h}$. Its marginalised probability density
function is shown in Figure \ref{q} as a histogram, with the median and
$1\sigma$ dispersion indicated in the legend. We also show the best-fitting
log-Normal distributions as solid lines, and detail their parameters and additional
distribution properties in Appendix \ref{D}. Two-parameter
marginalized distribution are shown for the axis ratio together with the
total baryonic mass $M_{\Star}$, and the local dark matter halo density
$\rho_{\rm h}\var{R_{\Sun}}$, in the two panels in Figure
\ref{corner_Mrho}, with the same color and legend codes as in Figure
\ref{corner}.

The stream generated by NGC 3201 does not help much constraining this parameter, giving a large error $q_\rho^{\rm h}=2.06\pm0.93$, with a rather asymmetric distribution that favors a prolate halo. This wide distribution is a consequence of the observed short section of the stream being located close to the pericentre, and with an equatorial projection that makes the stellar distribution insensitive to the variation of the axis ratio. Palomar 5 yields a much more powerful constraint of $q_\rho^{\rm h}=1.01\pm0.09$, implying the halo is rather close to spherical, without a dependence on the stellar mass. For M68 we obtain $q_\rho^{\rm h}=1.14^{+0.21}_{-0.14}$, in good agreement with Palomar 5 even though the streams explore very different regions of the gravitational potential. The M68 stream is compatible with a spherical halo but favoring a prolate one. For the M68 stream model, a more spherical halo is correlated with a lower stellar mass (see right panel of Figure \ref{corner_Mrho}), to compensate the acceleration on the stream (which has its best measured part passing 5 kpc above the disc) produced by each component. There is no significant correlation between the axis ratio and the other halo parameters for any of the streams. For Palomar 5, the main correlation appears with the heliocentric distance of the cluster which, at the same time, has a weak correlation with the proper motion of the cluster.

\subsection{Model with all streams together}\label{AllS}

 When all three streams are included together in the model, the halo
shape is better constrained using data over a larger region.
The M68 stream improves the constraint on the disc mass, which is larger
than the best fit for the other two streams (see Section \ref{sep}).
The three stream model results in
$M_{\rm d}^{\rm n} = (6.07\pm0.39)\pd{10}$ and
$M_{\Star} = (8.01\pm0.38)\pd{10}$ ${\rm M}_{\Sun}$, similar to the
results using only the M68 stream. This gives a vertical gravitational
acceleration at the solar radius $\Approx 1.8\sigma$ above the prior from
other observational constraints (eq.\ \ref{Kz}), and also a transverse
velocity of Sgr A$^*$ at $-1.8\sigma$ from the observation
(eq.\ \ref{SgrA}), similarly to the Palomar 5 case as discussed above.
The implied LSR velocity,
$\Theta_0 = 230.67\pm1.55$ ${\rm km}\,{\rm s}^{-1}$, and transverse
solar velocity
$\Theta_0 + V_{\Sun} = 244.38\pm0.91 \, {\rm km}\,{\rm s}^{-1}$, are
again consistent with other measurements as discussed in Section \ref{sep}. The rotation curve is consistent with observations, with slightly larger
velocities at large radius compared to single stream models. This is
related to the larger total dark halo mass of the three-stream model,
$M_{200}^{\rm h} = (1.08\pm0.22)\pd{12}$ ${\rm M}_{\Sun}$, $\Approx$ 14 per cent larger than the previous models.

The three stream model provides an improved constraint on the density
profile, with a similar conclusion of a flat inner profile, with
$\alpha=0.06\pm0.22$, within a large core of about $a_1=17^{+10}_{-3}$ kpc.
The core radius is nevertheless strongly correlated with the outer slope
$\beta=3.3^{+0.7}_{-0.3}$. The result for the axis ratio is $q_\rho^{\rm h} = 1.06\pm0.06$, again
consistent with spherical and slightly favoring a small deviation toward a prolate halo. As with all other streams separately, there is no significant correlation between the axis ratio and the other halo parameters. A larger baryonic mass, as preferred by the M68 stream, is what biases the axis ratio toward a more prolate halo.

\section{Discussion}\label{CTOS}

\subsection{Comparison to previous studies: observations}\label{comOBS}

 Several studies have been made of the Milky Way dark matter halo shape
using parametric models for the mass distribution constrained by
observational data, and comparing this to predictions from cosmological
simulations of Milky Way-like galaxies taking into account baryonic
effects. In this subsection, we review these studies focusing on the halo
axis ratio and compare them to our results.

\begin{figure*}
\includegraphics[width=1.0\columnwidth]{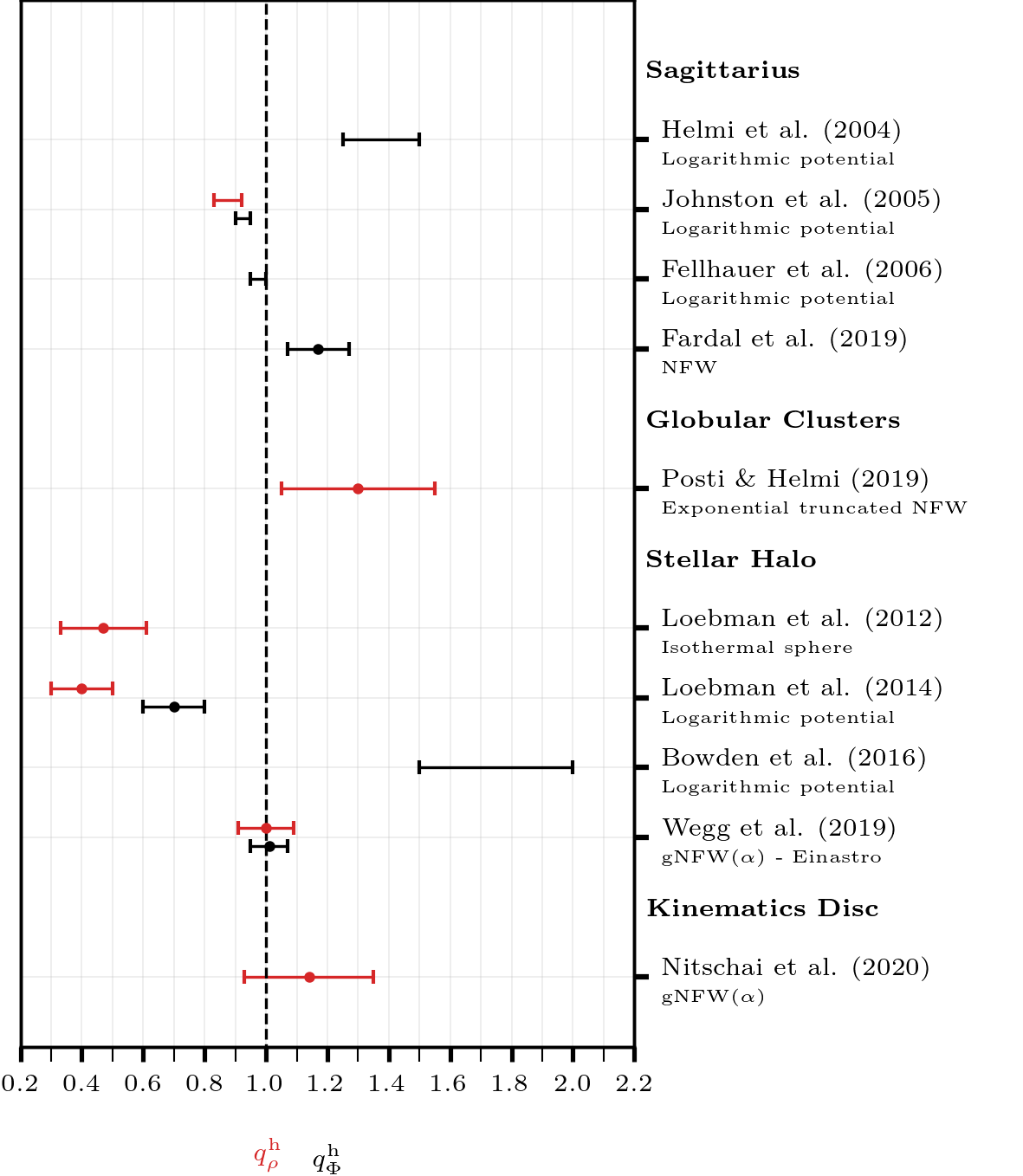}
\includegraphics[width=1.0\columnwidth]{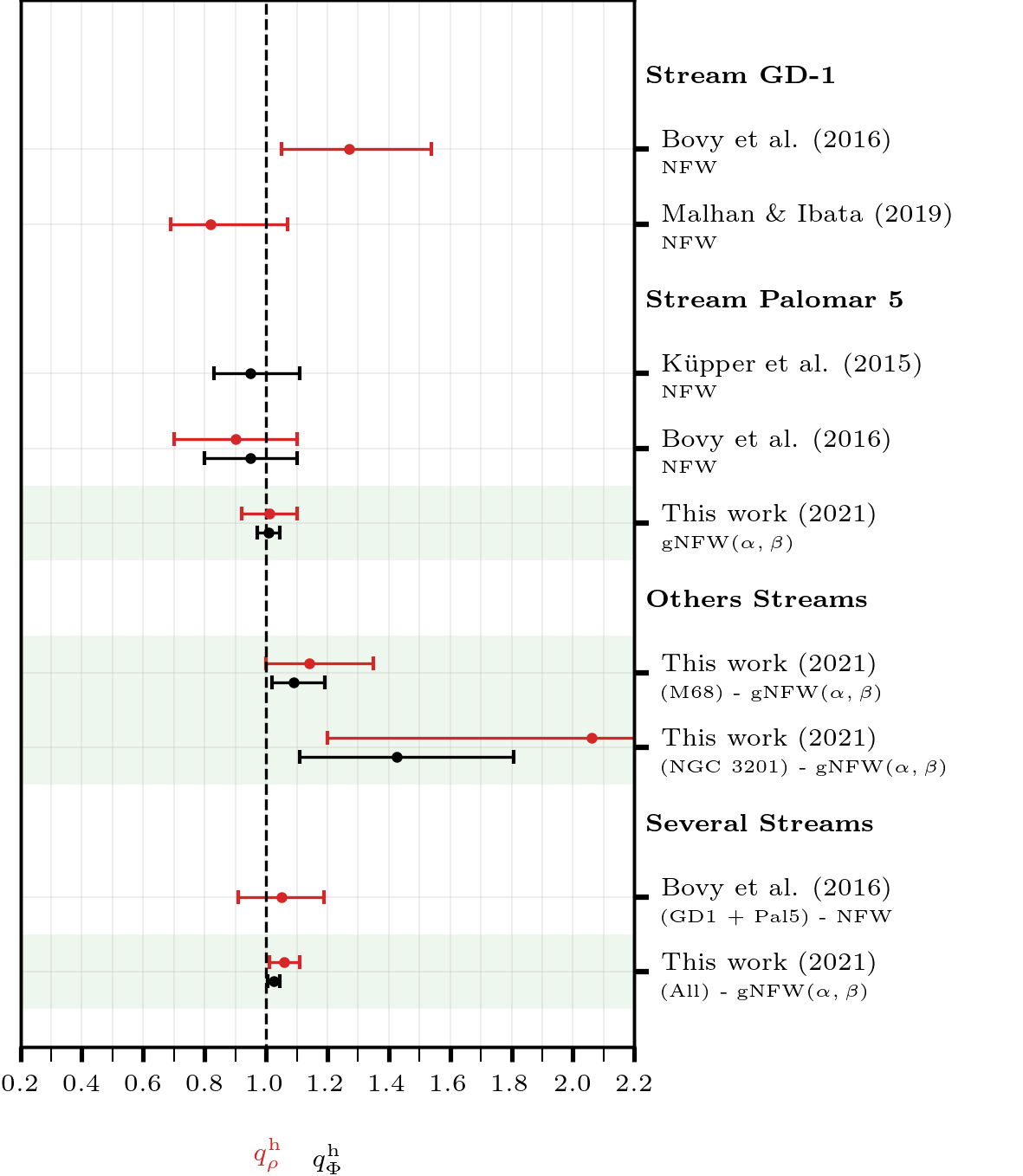}
\caption{A list of estimates of the axis ratio of the dark matter halo
density, $q_\rho^{\rm h}$ (red), and potential, $q_\Phi^{\rm h}$
(black), comparing our work and previous ones, grouped according to main
source of observational data (in boldface). The halo model used is
indicated under the reference, including the slope values for gNFW in
parentheses. Dots are mean or median value as stated in the source, and
bars are $1\sigma$ deviations (dots are absent when the result is
reported as an interval). Vertical dashed line marks the spherical case and green shaded regions
highlight our estimates.}
\label{comp}
\end{figure*}

 We consider studies of the Milky Way halo shape based on axisymmetric
analytic models consistent with dynamical equilibrium, with the symmetry
axis perpendicular to the Galactic disc. Many studies adopt a NFW density
profile for the halo, or a generalised version where the inner and outer
power-law slopes are free (gNFW). In other models, the halo potential is
assumed to follow the axisymmetric logarithmic potential,
$\Phi_h=\log[R^2+(z/q_\Phi^{\rm h})^2+r_c^2]$, with a core radius $r_c$.
Results from this type of studies are shown in Figure \ref{comp} for the
halo density axis ratio $q_\rho^{\rm h}$ (red) and the halo potential
axis ratio $q_\Phi^{\rm h}$ (black), with their quoted error bars.
The studies are grouped according to the main source of observational
data (in boldface), with the halo model that is used indicated under
each reference. The dashed vertical line indicates the spherical case ($q^{\rm h}=1$).

Early studies did not converge to a consistent picture. Some studies
using the Sagittarius stellar stream proposed triaxial shapes
\citep{2009ApJ...703L..67L, 2010ApJ...714..229L, 2013MNRAS.428..912D},
but were criticised for their instability and incompatibility with
constraints from Palomar 5 or Sagittarius's tidal streams
\citep[see e.g.][]{2013ApJ...765L..15I, 2013MNRAS.434.2971D, 2014MNRAS.437..116B, 2015ApJ...799...28P}, and we exclude them in Figure \ref{comp}. Using a sample of carbon stars, \citet{2001ApJ...551..294I} noticed that the Sagittarius stream is observed as a Great Circle, indicating that the dark halo is most likely nearly spherical at $16<r<60$ kpc. Also based on Sagittarius
stream, \citet{2004ApJ...610L..97H} obtained a prolate halo, while
\citet{2005ApJ...619..800J} and \citet{2006ApJ...651..167F} obtained a
shape much closer to spherical and slightly oblate. Likewise, using 
equilibrium models of halo stars,
\citet{2012ApJ...758L..23L, 2014ApJ...794..151L} found an oblate dark
matter halo while \citet{2016MNRAS.460..329B} obtained a prolate one.

 On the other hand, recent studies offer a more consistent picture,
indicating a nearly spherical halo, or a slightly prolate shape.
Using the Sagittarius stream mapped with RR Lyrae from Pan-STARRS1,
\citet{2019MNRAS.483.4724F} found $q^{\rm h}_\Phi=1.17\pm0.1$. This
is not far from the result of \citet{2019MNRAS.485.3296W} using RR Lyrae
halo stars in the radius range $r\Approx2$ to $20$ kpc, who conclude
that the halo is spherical with $q^{\rm h}_\rho=1\pm0.09$ (these
authors obtain the same result assuming a gNFW or a Einastro halo radial
profile). The study of \citet{2021MNRAS.508.5468H} assumes an
equilibrium distribution function of globular clusters to infer the halo axis
ratio. The distribution is computed in angle-action framework using the \textsc{Agama} package \citep{2019MNRAS.482.1525V} which is limited to spherical-oblate axisymmetric potentials. They found $0.963<q^{\rm h}_\rho$, strongly disfavouring a flattened dark matter halo. We note also that \citet{2020IAUS..353...96H} favor a halo axis ratio near $q_\rho^{\rm h} \simeq 1.5$ using a hypervelocity star and
assuming that it was ejected from the Galactic centre. In addition,
\citet{2020MNRAS.494.6001N} use disc kinematic data at
$R\Approx4$ to 12 kpc and a vertical height $|z| \Approx 2\,
{\rm kpc}$, favoring also a slightly prolate halo with
$q^{\rm h}_\rho=1.14\pm0.21$, although their error bar is also large.

In general, studies that have used stellar streams are all consistent
with each other. Their results are plotted in the right-hand panel of
Figure \ref{comp}, along with our estimates highlighted with a light
green shade. We do not include the first studies using the GD-1 stellar
stream \citep{2010ApJ...712..260K, 2015MNRAS.449.1391B} because they did
not constrain the dark matter halo directly but the overall potential of
the Galaxy. Using the GD-1 stream, \citet{2016ApJ...833...31B} find
$q_{\rho}^{\rm h} \simeq 1.27\pm0.27$, but \citet{2019MNRAS.486.2995M}
obtain $q_{\rho}^{\rm h} = 0.82^{+0.25}_{-0.13}$. The latter study uses
better constraints from a larger number of stars from the GDR2
catalogue; nevertheless, the results still have a large error bar and
are both compatible with a spherical halo.

\subsection{Comparison to our study based on stellar streams}

 We compare now these results based on the GD-1 stream with our models
of the M68 and NGC 3201 streams. The observed sections of the GD-1 and
M68 streams are at similar distances above the Galactic disc, and even
though the M68 stellar stream is $\Approx$ 5 kpc closer to the Galactic
centre than GD-1, they are still sensitive to a similar radial range of
the dark halo shape. Our estimate favours a prolate halo but is
compatible with a spherical shape, in better agreement with
\citet{2016ApJ...833...31B} but not incompatible with
\citet{2019MNRAS.486.2995M}. In the case of the NGC 3201 stream, located at
similar distance from the Galactic centre as GD-1 but closer to the
disc plane, our error bar using only NGC 3201 is very large but still
favours a prolate halo, compatible with M68 and
\citet{2016ApJ...833...31B}.

 The case of the Palomar 5 stream is particularly interesting, because
its position, far above the disc at a larger distance from the Galactic
centre, makes it a better probe of the dark halo shape.
\citet{2015ApJ...803...80K} carried out a study using sky coordinates
and line-of-sight velocities of several members of this stream. Modeling the
Milky Way with a Miyamoto-Nagai disc potential and a NFW halo density
profile, and using a Bayesian framework developed by
\citet{2014ApJ...795...94B}, they infer $q_\Phi^{\rm h}=0.95^{+0.16}_{-0.12}$. At the same time,
\citet{2016ApJ...833...31B} use a similar model and data but a different
stream-fitting methodology based on action-angle modelling introduced in
\citet{2014ApJ...795...95B}, obtaining $q_\rho^{\rm h}=0.9\pm0.2$. The latter authors also combine the Palomar 5 stream with GD-1 to obtain
the improved constraint $q_\Phi^{\rm h}=1.05\pm0.14$.

 These estimates agree within the quoted observational errors, and they
are also compatible with our result for Palomar 5 alone, 
$q_\rho^{\rm h}=1.01\pm0.09$. Our error bar is smaller, even though our
halo model has more free parameters. The likely reason is that we have a
larger sample of stars with five phase-space parameters measured by GDR2
and 15 stars with radial velocity. We conclude that our measurements are
fully consistent with these previous studies, including our combined
result from the three streams we use, $q_\rho^{\rm h}=1.06\pm0.06$. All of them favour a halo that is close to spherical, eliminating in particular the possibility of a highly oblate halo. This conclusion applies to the range of radii probed by these streams,
$10 \LessSim r \LessSim 20\, {\rm kpc}$.

\subsection{Predictions from cosmological simulations}\label{comCS}

 Cosmological simulations including only dark matter predict that the
gravitational evolution of random initial fluctuations should lead to highly triaxial halos. However, when models of the behaviour of the baryonic
components are included, with the complexities of disc and bar formation
near the centre, halos are found to generally become more rounded owing
to the accumulation of a central mass dominated by baryons. In fact,
the potential of a triaxial halo is generally supported by highly
populated box orbits that are aligned along the long axis of the
potential. When a concentrated structure grows at the centre, dark
matter particles in initially box orbits can be scattered in random
directions when passing close to the centre. As a galaxy forms and grows in mass, this mechanism can make the dark matter distribution increasingly spherical in the inner regions of the halo.

\begin{figure}
\includegraphics[width=1.0\columnwidth]{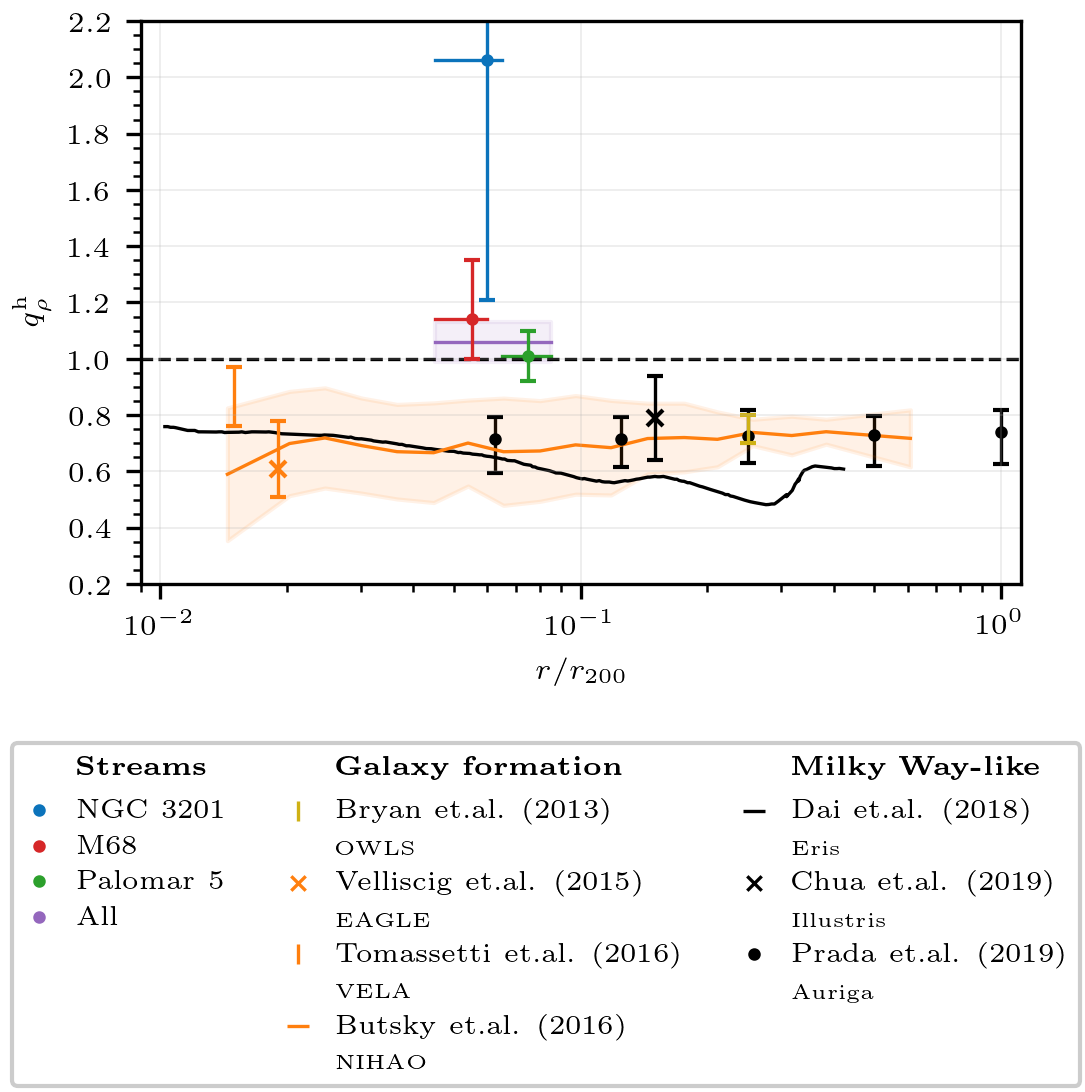}
\caption{Comparison of our estimates of $q_\rho^{\rm h}$ from each of
our three streams (red for M68, blue for NGC 3201, green for Palomar 5),
and all streams together (purple), to predictions of the halo axis ratio
from simulations of galaxy formation. Black dots with error bars and
shaded areas are mean values and $1\sigma$ ranges in the halo axis
ratios measured in disc galaxy host halos along the axis perpendicular
to the disc plane (no dot is shown when the result is given as an
interval in the referenced works). The simulation name appears under each reference. The simulated galaxies
have $r_{200}\Approx\Range{200}{240}$ kpc. Our stream model results are
shown at the mean distance of each stream, with its radial range shown as
a horizontal bar, and the whole range indicated by the all streams model
as the purple line. The dashed horizontal line marks the spherical shape.}
\label{comp_sim}
\end{figure}

Results from cosmological simulations of galaxy formation generally
agree that the majority of disc galaxies end up with dark matter halos 
that are oblate in the inner regions, with the short axis aligned close
to perpendicular to the disc plane
\citep[e.g.,][]{2005ApJ...627L..17B,2016MNRAS.460.3772S,2019MNRAS.490.4877P}. In Figure \ref{comp_sim} we plot estimates of $q_\rho^{\rm h}$ in galactic
halos resembling the Milky Way at $z=0$, with a total halo mass close to
$10^{12}\, M_{\odot}$, obtained from numerical simulations of galaxy
formation. The axis ratios are measured as a function of the distance to
the centre of the halo hosting the galaxies, assuming the short axis to
be perpendicular to the plane of the disc galaxy in the simulation.
The results are from a variety of galaxy formation simulations using
cosmological initial conditions, in
\cite{2013MNRAS.429.3316B, 2015MNRAS.453..721V, 2016MNRAS.458.4477T, 2016MNRAS.462..663B, 2018ApJ...858...73D, 2019MNRAS.484..476C, 2019MNRAS.490.4877P}. In comparison, our inferred values for the halo axis ratio from each
individual streams and the three streams together are shown as dots with
our usual colour code, with the error bars in $q_\rho^{\rm h}$, and a
radial value and range indicating the Galactocentric radius of the stream
section that is observed in each case. For the model of all streams
together, the error is shown as the shaded purple area and the radial
range is for all three streams.

 All simulations predict oblate dark halos. Taking the estimates in
the radial range $r\Approx\Range{5}{20}$ kpc
(or $r\Approx \Range{0.02}{0.1}\, r_{200}$), to which our observational
constraints from the stellar streams we use are sensitive to, the
simulations predict $q_\rho^{\rm h}\Approx0.74\pm0.15$. These results
that the majority of disc galaxies should be surrounded by oblate halos.
Taking the $1\sigma$ dispersion from these simulations, we find that
our estimate for the Milky Way galaxy axis ratio is discrepant from
this prediction by about $2\sigma$, with the error being dominated by
the range in the axis ratio of simulated galaxies rather than our
observational determination.  We therefore conclude that if the results of these numerical simulations
are correct, our estimate for the Milky Way halo axis ratio would imply
that the Milky Way galaxy is an anomalous one, being a rare case where
the halo has a nearly spherical or slightly triaxial shape, instead of
the average oblate halo with axis ratio
$q_\rho^{\rm h}\Approx0.75$ we should expect for a typical galaxy.

\subsection{Influence of the Magellanic Clouds}\label{comMC}

 An important question in relation to the observationally inferred
estimates of the shape of the Milky Way dark matter halo, and the
comparison to predictions from numerical simulations, is the influence
that the Magellanic Clouds may have in distorting this halo shape in
their encounter with the Milky Way galaxy. Recent studies have shown
that the Magellanic Clouds are in their first orbital passage through
their pericentre around the Milky Way galaxy, and that their associated
dark matter halo may be as massive as $\Approx 1/4$ of the Milky Way dark
matter halo \citep[see e.g.][]{2019MNRAS.487.2685E, 2020ApJ...890..110G, 2020ApJ...893..121P}.
Thus, the merger of the Milky Way with the Magellanic
Clouds that is unfolding at present is not so much a "minor merger",
but a merger of two galactic systems that are more closely comparable
in mass than was thought in the past.

The Milky Way dark matter halo at a distance $r$ from the centre is
in a dynamical state governed by the acceleration $g_{\:\!\MW}\Approx
GM_{\MW}/r^2$, where $M_{\MW}$ is the total Milky Way mass within $r$,
and is perturbed by the gravitational tide of the Magellanic Cloud
system at a distance $d$ from the Milky Way centre. This tidal
acceleration is, to first order in $r/d$, $g_{\:\!\MC}\Approx G M_{\MC}\, r/d^3$, so the ratio of the two
accelerations is:
\begin{equation}
 \dfrac{g_{\:\!\MW}}{g_{\:\!\MC}}  = \dfrac{M_{\MW}\, d^3}{M_{\MC}\,r^3} ~.
\end{equation}
Our measurements of the Milky Way halo shape are at $r < 20 \, {\rm kpc}$, and the Galactocentric distance to the Magellanic clouds is $d\simeq 50 \, {\rm kpc}$, so we conclude that for $ M_{\MW}/ M_{\MC} \simeq 4$, the tidal influence of the LMC is no larger than $\Approx$ 2 per cent of the usual acceleration in the Milky Way for the stellar streams we study. In addition, the distance of the streams to the LMC is always greater than about 40 kpc, with the closest approximation being approximately 45 kpc for NGC 3201, 37 kpc for M68, and 40 kpc for Palomar 5. We therefore conclude that the Magellanic Clouds should not be affecting our conclusions, although it is certainly important to include their effect for studies going to larger radius or seeking higher accuracy in the halo shape determination.

\subsection{Consequences for the Milky Way halo dynamical equilibrium state}
\label{comHDE}

A spherical dark matter halo surrounding the Milky Way galaxy in the presence
of the disk cannot have an isotropic velocity dispersion. In order to be
supported in the oblate gravitational potential that results from the combined
mass distribution of the halo and disk, the velocity dispersion must be
higher in the vertical direction (perpendicular to the disk) compared to the
two directions in the disk plane. This is a consequence of the tensor virial
theorem and the Vlasov equations of dynamical equilibrium \citep{2008gady.book.....B}. The velocity
anisotropy is important if the halo is substantially less oblate than the
isopotential surfaces, in the region where the disk mass is comparable to the
halo one. We have found the Milky Way halo to be close to spherical (or
slightly prolate) in the radial interval of $10$ to $20$ kpc, which suggests
the presence of this anisotropic velocity dispersion of the dark matter in
the Milky Way.

This anisotropy in the velocity dispersion will need to be further analyzed
in future work to quantify its presence, but if real, it would have to
originate from an originally prolate halo with a long axis perpendicular to
the disk formed in the assembly process of the Milky Way halo. In addition, interactions of the dark matter particles with the baryonic components of the
Milky Way (in particular the bulge density cusp and a rotating bar, which result
in random scatterings of distant particles moving through the central galaxy region)
should tend to isotropize the dark matter velocity dispersion and thereby increase
the halo oblateness. This probably requires a more strongly prolate shape of the
original Milky Way halo shape to reach a nearly spherical configuration in the
present Galaxy at radii in 10 to 20 kpc range.

Future studies will need to address this issue of the required
initial anisotropic configuration of the halo to support the halo
shape at the radii where the baryonic contribution to the potential
is important.

\section{Conclusion}\label{con}

 Stellar streams provide us with a powerful methodology to measure the
gravitational potential of the Milky Way and to infer the distribution
of mass which, when taking into account the contribution of visible
baryonic matter, can give us indications on the distribution of dark
matter. One of the most interesting constraints we can derive is the
departure from sphericity of the dark matter halo, and test if the halo
is oblate or prolate with respect to the disc at different radii. The
distribution of the halo axis ratio at different radii can be
compared with predictions of galaxy formation from cosmological
simulations. In the past, the mass distribution could be constrained only
from the kinematic distributions of various tracers in the Milky Way using
assumptions of dynamical equilibrium, but stellar streams from tidally
disrupted systems allow an indirect measurement of accelerations, because
the stream trajectory indicates the orbit of the tidally truncated system,
except for small deviations that can be modelled and corrected \citep[see e.g.][]{2015ApJ...803...80K,2016ApJ...833...31B, 2019MNRAS.486.2995M}.

 We have used three streams to model the Milky Way potential in this
paper, arising from the tidal stripping of globular clusters NGC 3201, M68, and Palomar 5. We expect that in the future, the large number
of other streams being discovered will be used in conjunction to obtain
the best constraints on the Milky Way potential, but this paper is our
initial attempt to obtain such constraints based on three streams that
appear particularly interesting at this time due to their proximity and
available members in the \textit{Gaia} catalogue. After selecting a list of
members of these streams with our maximum likelihood method, we have
fitted a model of the Galactic potential based on 5 free parameters
of an axisymmetric dark matter halo (mass, inner and outer slope, core
radius, and axis ratio), while adding other free parameters for the
baryonic components that are subject to various prior observational
constraints (the Sun's position and velocity, the rotation curve in the
radial range from 5 to 25 kpc, other star kinematics, and velocities of
distant Milky Way satellites).

To show how the constraints arise from each stream, our results have been presented as parameterized models fitted to each
stream individually, and to all three stream together. Our interest focuses mainly on the
dark matter halo oblateness, to use this as a test of dark matter
theories that can predict the distribution of the axis ratio. We find
that while the NGC 3201 stream is not very sensitive to this axis ratio,
the Palomar 5 stream gives a strong constraint of
$q_\rho^{\rm h}=1.01\pm0.09$, and the M68 stream yields
$q_\rho^{\rm h}=1.14^{+0.21}_{-0.14}$, owing to favorable trajectories
of these streams that are sensitive to the acceleration differences
introduced by the halo oblateness. The parameter degeneracy is reduced by the
priors from other available data on the rotation curve and vertical
velocity dispersion of the disc. In the case of the M68 model, the oblateness parameter is correlated
mainly with the disc mass. A more massive disc is preferred, but the final constraint on
$q_\rho^{\rm h}$ is compatible with the other streams. Our combined result on the axis ratio from the three streams is
$q_\rho^{\rm h}=1.06\pm0.06$, consistent with a spherical halo with
a statistical preference for a slightly prolate halo. Our model assumes
a halo axis ratio that is independent of radius, and these constraints
are to be understood as applying near the radius that is probed by the
streams, at $r \Approx 10$ to $20\, {\rm kpc}$.
This result agrees with previous studies using different observational
data and fitting methodology.

 Our best fit model also demands a very shallow density profile for the
dark matter halo, with inner slope $\alpha$ close to zero and a large
core radius of $\Approx 15\, {\rm kpc}$. The flatness of the density profile
in the inner region is also interesting to test the way that the
formation of the disc and bar and the presence of gas inflows and
outflows over the history of the Milky Way may have flattened the
central parts of dark matter halo. Nevertheless, the inner dark matter
distribution is probably degenerate with the baryonic mass component in
the inner disc, bar and bulge. Our model simply includes an exponential
disc with no inner cutoff and a bulge with only one free parameter (the
mass), and a more careful treatment of the mass distribution at
$r \LessSim 5 \, {\rm kpc}$ is needed to more rigorously test
constraints on the inner dark matter profile. Our model constraints on
the outer dark matter density slope and total mass $M_{\rm 200}$ are also
mainly dependent on the constraints from external satellite kinematic
data we use.

 The Milky Way dark matter halo density model should be greatly improved
in the future by including the large number of stellar streams that are
being discovered with a wide range of orbits in the Milky Way halo. Some
of the most interesting cases are the streams generated by globular
clusters NGC 5466 and M5, with similar characteristics and locations as
the streams used in this study. A greater variety of models and
parameters should also be included, and the impact of the gravitational
perturbation by the Magellanic Clouds and other massive satellites
should be incorporated as we probe the halo density profile and
oblateness at larger radius and/or with greater accuracy than in
this study.

 At present, we can already say that most of the cosmological simulations
that have been analyzed in relation to the question of the oblateness of
galactic halos seem to predict oblate halos, with axis ratio lower than
the 2-$\sigma$ lower limit from our study. This may be a possible
discrepancy with Cold Dark Matter theories, indicating that either the
dark matter has some new property that tends to make halos more
spherical in the inner parts, or that the Milky Way is a peculiar galaxy
with its halo long axis perpendicular to the disc, while the results of
simulations indicate that most galaxies should have the oblate halos.
We have pointed out that a spherical halo, in the presence of the
gravitational potential of the disc, actually needs to maintain an
anisotropic velocity dispersion, with greater dispersion along the
vertical axis compared to the horizontal ones, to maintain its spherical
shape in equilibrium, and this becomes important at the radius where the
disc and halo contributions to the gravitational potential are
comparable, at $r\Approx 10\, {\rm kpc}$. This suggests that it is
difficult to avoid having an oblate dark matter halo in the inner
regions of the galaxy, if random scatterings (caused, for example, by
a rotating bar or the central density cusp and black hole in the bulge)
tend to isotropize the orbital motions of the dark matter. Future
studies, using improved data from streams and stellar kinematic
constraints, and more general models for the gravitational potential,
will hopefully clarify these questions on the Milky Way dark matter
halo.


\section*{Acknowledgements}
It is a pleasure to thank John Magorrian for helpful comments and discussions on this paper, and the Oxford Galactic Dynamics Group for their help and support. We are also grateful to the anonymous referee for careful reading and suggestions for improving this paper.

This work received support from the Spanish Maria de Maeztu grants
CEX2019-000918-M and MDM-2014-0369 to the Institut de Ciències del Cosmos - Universitat de Barcelona (ICCUB), and the grant
PID2019-108122GB-C32. This work has made use of data from the European Space Agency (ESA) mission
\textit{Gaia} (\url{https://www.cosmos.esa.int/gaia}), processed by the
\textit{Gaia} Data Processing and Analysis Consortium
(DPAC, \url{https://www.cosmos.esa.int/web/gaia/dpac/consortium}). Funding
for the DPAC has been provided by national institutions, in particular, the
institutions participating in the \textit{Gaia} Multilateral Agreement.

This research used the Python packages: \textsc{Astropy} \citep{2022ApJ...935..167A}, \textsc{Matplotlib} \citep{Hunter:2007}, \textsc{mpmath} \citep{2017zndo...1476881J}, \textsc{Numpy} \citep{harris2020array}, \textsc{SciPy} \citep{2020SciPy-NMeth}.


\section*{Data Availability}

The data on which this article is based are available in the article itself and in the references included therein.




\bibliographystyle{mnras}
\bibliography{bib/ref.bib}



\clearpage
\appendix


\section{Observed sections of NGC 3201 and M68 streams}\label{AppA}

\begin{figure}
\includegraphics[width=1.0\columnwidth]{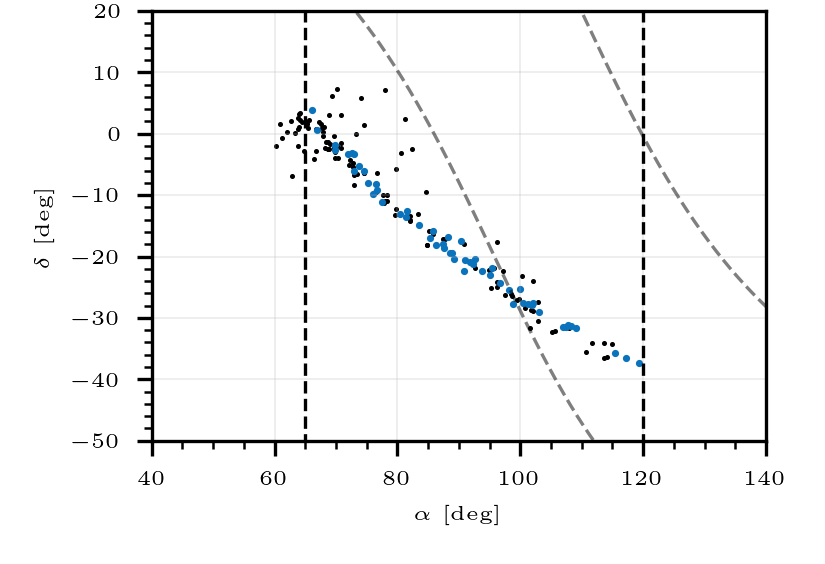}\\
\includegraphics[width=1.0\columnwidth]{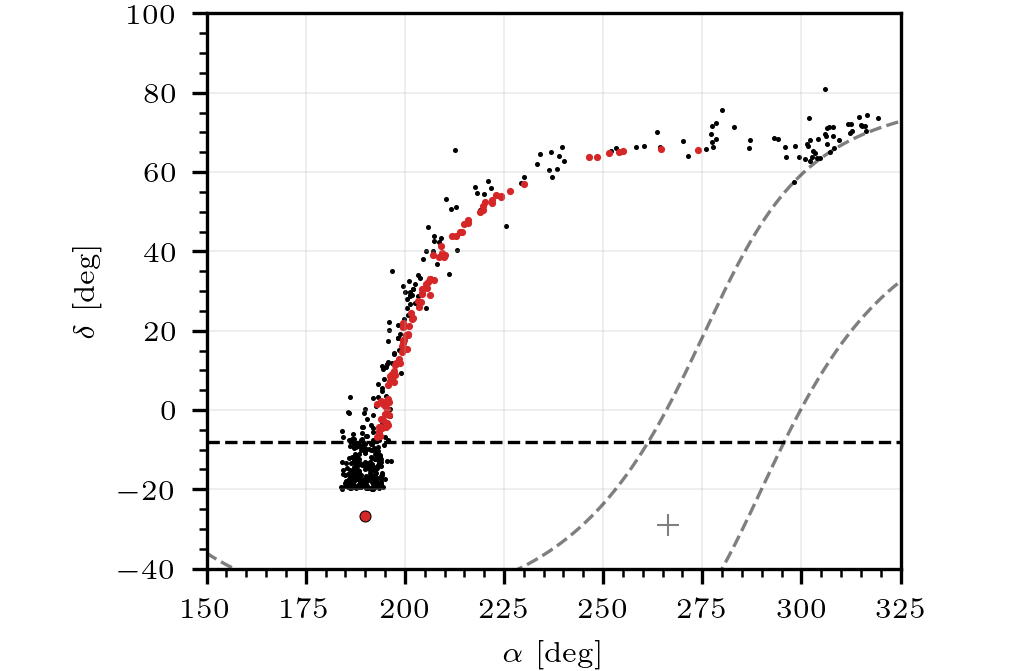}
\caption{Sky map in equatorial coordinates of a sample of pre-selected stars (black dots) and the final selection of stream stars (coloured dots) used to constrain the potential of the Milky Way. Grey dashed lines indicate a Galactic latitude $b = \pm15$ deg and the grey cross the Galactic centre. \textit{Top panel:} Blue dots mark the selected NGC 3201 stream stars. The black dashed lines at $\alpha=65$ and $120$ deg mark the limits of the selection zone. \textit{Bottom panel:} Red dots mark the selected M68 stream stars. The black dashed line at $\delta=-8$ deg marks the boundary of the selection zone. The large red dot marks the current position of the cluster.}
\label{sec_stream}
\end{figure}

The section of the NGC 3201 used to constrain the Galactic potential is populated by 54 stars selected in \citetalias{2021MNRAS.504.2727P} within the limits of $65<\alpha<120$ deg and located between about 3 and 4 kpc from the Sun. In the top panel of Figure \ref{sec_stream} we show the selected stars in blue, and in black, the foreground passing the pre-section cut defined in Section 3.3 of \citetalias{2019MNRAS.488.1535P} with a threshold $P_\SM{REG} = 0.5$ ${\rm yr}^3 \, {\rm deg}^{-2} \, {\rm pc}^{-1} \, {\rm mas}^{-3}$. We also mark the boundaries of our selection with vertical dashed lines.

In summary, the pre-section removes stars with a parallax $\pi > 1/0.3$ mas and selects the stars compatible with the HR-diagram of NGC 3201 using the method described in Section 3.3 of \citetalias{2021MNRAS.504.2727P}. It also selects stars with proper motions $\GtrSim 10$ mas yr$^{-1}$ in a region around the orbit of the cluster wide enough not to eliminate potential candidate stars. For example, we can see how stars located far from the stream, at $\alpha\Approx 80$ and $\delta\Approx6$ deg pass through these pre-selection cuts. The pre-selection is more restrictive in the colour and magnitude space, where we exclude stars with large observational uncertainties. This does not alter the stream track, but enhances the star overdensity in sky coordinates.

Within the defined limits, the stream can be seen without ambiguity due to its high population and low number of foreground stars (in Section 3.1 of \citetalias{2021MNRAS.504.2727P} we show how the expected number of foreground stars along the overdensity is between 4 and 8). The final selection consists of stars that intersect a density model of the stream. We compute the intersection using the method described in Section 4.3 of \citetalias{2019MNRAS.488.1535P}. The density model of the stream is approximately a cylinder that follows the overdensity of stars in phase-space. Its diameter is determined by a selection threshold $\chi_{\rm sel}$ given in Section 3.4 of \citetalias{2021MNRAS.504.2727P}, optimised to include stars along the overdensity but exclude the observed foreground. Some stars located along the stream have not been included in the final selection. This is due to their large proper motion uncertainties compared to the selected stars. Their exclusion from our sample does not change the track of the stream, and therefore the constraints on the Galactic potential obtained with this selection.

The stream of M68 is shown in red in the bottom panel of Figure \ref{sec_stream}. We select 98 stars from the final selection of \citetalias{2019MNRAS.488.1535P}. We also plot the foreground stars which pass through a similar pre-selection cut as described above. As in the previous case, the pre-selection cut does not remove many stream stars using their position in phase-space. We note that some stars located at about 25 deg perpendicular to the orbit pass through these cuts. The most restrictive cuts are applied in colour and magnitude space, as the loss of faint stars with large observational uncertainties does not alter the stream track.

The stream is clearly visible when it is close to the Sun, at about 5 kpc, in the interval $190<\alpha<260$ deg. We have excluded the region with $\delta < -8$ deg because the stream is located at about 10 kpc from the Sun and beyond, and appears to be projected closer to the disc. In this region, there is a high level of foreground star contamination. For similar reasons, no stars have been selected for $\alpha > 275$ deg.

There are several stars located around the stream, especially above our selection, between $190<\alpha<240$ deg. These stars form a diffuse envelope or cocoon, similar to the presented by other streams such as GD-1 \citep{2019ApJ...881..106M} or Jhelum \citep{2023AandA...669A.102W}. These stars can be separated from the stream because their proper motions are significantly larger, and they are aligned in proper motion space following a different orbit clearly distinguishable from the main stream. This orbit cannot be followed by either the cluster or its stream for any reasonable Galactic potential. This may be a consequence of the stellar envelope being closer to the Sun than the stream. A detailed analysis of the structure of the stream and its diffuse envelope will be realised when radial velocities and better measurements are available from future releases of the \textit{Gaia} catalogue.

Similarly to the previous case, several stars along the stream have not been selected due to their comparatively large proper motion uncertainties. It is likely that the majority of these stars belong to the stream. In Section 4.3 of \citetalias{2019MNRAS.488.1535P} we showed that we expect only 1 or 2 foreground stars along the stream overdensity within the defined limits. Excluding these stars does not change the stream track, and therefore the constraints on the potential obtained by fitting this star selection.

\section{Final selection of Palomar 5 tidal stream members}\label{E}

To select the stars most likely to belong to the Palomar 5 stellar stream, we use the method described in \citetalias{2019MNRAS.488.1535P}. First, we apply the pre-selection cuts defined in Section 3.3 of \citetalias{2019MNRAS.488.1535P} to reduce the number of foreground stars surrounding the stream. These cuts basically select stars near the orbit of the cluster, in a $\pm 20$ Myr section of the orbit from the cluster centre. They also remove the stars belonging to the globular cluster Palomar 5 and M5 to avoid detecting overdensities that do not correspond to streams. After the pre-selection, we obtain 320 302 GDR2 sources. We apply the same pre-selection to a simulation of the GDR2 catalogue, the 18th version of the Gaia Object Generator \citep[GOG18,][]{2014AandA...566A.119L} obtaining 450 622 sources. This 30 per cent difference can be explained by imperfect modeling of the stellar halo in GOG18, by inaccuracies in the simulation of GDR2 uncertainties, or because GDR2 does not include all sources with $\Gband$ magnitude $G<21$ mag in low exposure areas.

We apply the maximum likelihood method explained in Section 2 of \citetalias{2019MNRAS.488.1535P} to compute the best-fitting parameters of the stream model, as well as the statistic $\Lambda$ indicating the confidence level with which the stream is detected. When $\Lambda > 6.6$, the existence of the stream is confirmed at the 99 per cent confidence level, as opposed to the null hypothesis that no stream is present in the pre-selection. We detect the stellar stream with $\Lambda = 14.44$, which implies a high statistical significance of the detection. We compute an accurate phase-space density model of the stream using the best-fitting configuration of the free parameters. We select the stars with the largest intersection with this model. We define a threshold for the value of the intersection $\chi_{\rm sel}$, and choose stars with $\chi_{\rm sel} > 4.6$ yr$^3$ deg$^{-2}$ pc$^{-1}$ mas$^{-3}$. We obtain 229 stars from the GRD2 catalogue compatible with the phase-space density model of the stream. For the chosen threshold, we select no stars from the GOG18 catalogue. This minimizes the number of expected foreground stars erroneously selected as Palomar 5 stream members.

Finally, we only select stars that are compatible in colour and magnitude with the H-R diagram of Palomar 5. We follow the procedure described in Appendix D of \citetalias{2019MNRAS.488.1535P} and include the correction for dust extinction described in Appendix B of \citetalias{2021MNRAS.504.2727P}. In Table \ref{sel0}, we list the 126 star candidates belonging to the Palomar 5 tidal stream selected from the GDR2 catalogue. None of these stars have radial velocity measured by \textit{Gaia}, but 15 of them match stars with radial velocity measured by \citet{2017ApJ...842..120I}. We list their values in Table \ref{rv_table}.

\begin{table*}
\caption[]{\small{Stars compatible with the best-fitting phase-space density model of the tidal stream of Palomar 5 and
its H-R diagram from GDR2 after dust extinction correction.}}
\begin{center}
\begin{tabular}{rrrrrrrrrr}
\toprule

\multicolumn{1}{c}{N}
&\multicolumn{1}{c}{source\_id}
&\multicolumn{1}{c}{$\pi$}
&\multicolumn{1}{c}{$\delta$}
&\multicolumn{1}{c}{$\alpha$}
&\multicolumn{1}{c}{$\mu_\delta$}
&\multicolumn{1}{c}{$\mu_{\alpha*}$}
&\multicolumn{1}{c}{\scalebox{0.8}{$G_{\rm BP}\!-\!G_{\rm RP}$}}
&\multicolumn{1}{c}{$G$}
&\multicolumn{1}{c}{$\chi_{\rm sel}$}\\
&
&\multicolumn{1}{c}{\units{mas}}
&\multicolumn{1}{c}{\units{deg}}
&\multicolumn{1}{c}{\units{deg}}
&\multicolumn{1}{c}{\units{mas yr$^{-1}$}}
&\multicolumn{1}{c}{\units{mas yr$^{-1}$}}
&\multicolumn{1}{c}{\units{mag}}
&\multicolumn{1}{c}{\units{mag}}
&\multicolumn{1}{c}{\units{\scalebox{0.8}{yr$^{3}$ deg$^{-2}$ pc$^{-1}$ mas$^{-3}$}}}\\

\midrule

1&6327240546525053824&$-0.1349$&$-8.1333$&$222.5158$&$-2.4483$&$-2.4054$&$1.0417$&$17.5562$&$5.9454\text{E}\:\!\:\!\plus\:\!00$\\
2&6327454916932476800&$-0.0426$&$-7.9192$&$222.6962$&$-2.4659$&$-2.3107$&$1.0067$&$17.8507$&$8.2674\text{E}\:\!\:\!\plus\:\!00$\\
3&6333632175119962368&$0.1395$&$-6.8635$&$223.3798$&$-2.4777$&$-2.6496$&$0.9335$&$19.4622$&$5.6500\text{E}\:\!\:\!\plus\:\!00$\\
4&6333638840909351808&$0.3333$&$-6.7025$&$223.7207$&$-2.6044$&$-2.5148$&$1.0312$&$17.8252$&$1.3563\text{E}\:\!\:\!\plus\:\!01$\\
5&6334133694157081856&$0.0026$&$-5.8463$&$223.8334$&$-2.3926$&$-2.7662$&$1.1181$&$17.5348$&$1.2756\text{E}\:\!\:\!\plus\:\!01$\\
6&6337139376694381440&$-0.1242$&$-5.6805$&$223.9155$&$-2.4263$&$-2.7577$&$0.9366$&$17.4458$&$4.8662\text{E}\:\!\:\!\plus\:\!00$\\
7&6334151110248964224&$0.0510$&$-5.6672$&$224.4260$&$-3.1251$&$-2.0798$&$1.0594$&$18.7005$&$1.4053\text{E}\:\!\:\!\plus\:\!01$\\
8&6334300162794237184&$0.5529$&$-5.0987$&$224.7233$&$-2.7727$&$-1.9366$&$1.0851$&$19.0335$&$6.7220\text{E}\:\!\:\!\plus\:\!00$\\
9&6338851453738047488&$0.1009$&$-4.2145$&$224.9720$&$-2.6403$&$-2.1383$&$1.0813$&$18.4176$&$9.9368\text{E}\:\!\:\!\plus\:\!00$\\
10&6334298169929419392&$0.1771$&$-5.0869$&$224.9823$&$-2.4609$&$-2.7599$&$1.0623$&$18.1751$&$6.6524\text{E}\:\!\:\!\plus\:\!00$\\
11&6338874990158885376&$0.1463$&$-4.1744$&$225.0625$&$-2.7762$&$-2.3852$&$1.0872$&$17.7758$&$3.4422\text{E}\:\!\:\!\plus\:\!01$\\
12&6338879564298848640&$-0.0324$&$-4.0963$&$225.3011$&$-2.4097$&$-1.2524$&$1.1311$&$18.9155$&$4.8295\text{E}\:\!\:\!\plus\:\!00$\\
13&6338859562636251904&$0.2195$&$-4.2534$&$225.3258$&$-2.5176$&$-1.7848$&$1.1024$&$18.0456$&$3.3353\text{E}\:\!\:\!\plus\:\!01$\\
14&6338880487717174528&$0.0521$&$-4.0583$&$225.4573$&$-2.3537$&$-3.1365$&$1.1220$&$17.6939$&$1.0366\text{E}\:\!\:\!\plus\:\!01$\\
15&6338869149003524224&$-0.0925$&$-4.0479$&$225.5377$&$-3.0576$&$-2.0272$&$1.0265$&$19.2443$&$2.1193\text{E}\:\!\:\!\plus\:\!01$\\
16&6339016002525065472&$0.0639$&$-3.5563$&$225.7535$&$-2.3171$&$-3.4270$&$1.1729$&$19.0526$&$5.8216\text{E}\:\!\:\!\plus\:\!00$\\
17&6338987758819995776&$-0.1428$&$-3.6964$&$225.7778$&$-2.4940$&$-2.3050$&$1.1700$&$18.7510$&$8.4701\text{E}\:\!\:\!\plus\:\!01$\\
18&6339017823591238400&$0.0527$&$-3.5147$&$225.8524$&$-2.6521$&$-3.2807$&$1.3766$&$20.0483$&$7.4983\text{E}\:\!\:\!\plus\:\!00$\\
19&6339403580374320896&$0.2071$&$-3.2162$&$226.1388$&$-2.8443$&$-2.8832$&$1.1255$&$19.7879$&$5.9449\text{E}\:\!\:\!\plus\:\!00$\\
20&6339405195282047232&$0.3347$&$-3.1899$&$226.1702$&$-1.9488$&$-2.9943$&$1.1956$&$17.8007$&$7.7164\text{E}\:\!\:\!\plus\:\!00$\\
21&6339486112465206528&$0.2516$&$-2.9349$&$226.3743$&$-2.9628$&$-2.5859$&$1.1431$&$18.1520$&$2.8470\text{E}\:\!\:\!\plus\:\!01$\\
22&6339492091059869696&$0.1419$&$-2.7626$&$226.4066$&$-1.7571$&$-2.2625$&$1.0337$&$19.6753$&$5.6089\text{E}\:\!\:\!\plus\:\!00$\\
23&6339498589346000768&$-0.0427$&$-2.7155$&$226.4922$&$-2.9333$&$-2.4861$&$0.5547$&$17.4368$&$8.3411\text{E}\:\!\:\!\plus\:\!00$\\
24&6339607745939180928&$0.2627$&$-2.2781$&$226.9747$&$-1.9119$&$-2.3462$&$0.9989$&$19.5513$&$1.8167\text{E}\:\!\:\!\plus\:\!01$\\
25&6339642724153085056&$-0.4911$&$-1.9832$&$227.2439$&$-2.5559$&$-3.2388$&$0.7922$&$19.9143$&$5.0767\text{E}\:\!\:\!\plus\:\!00$\\
26&6339639666136263040&$-0.2089$&$-2.0898$&$227.2613$&$-2.4832$&$-2.1312$&$0.8322$&$17.1603$&$1.3918\text{E}\:\!\:\!\plus\:\!02$\\
27&4418107238191732352&$-0.0317$&$-1.9491$&$227.3442$&$-3.2361$&$-2.4779$&$0.7086$&$20.1749$&$5.3026\text{E}\:\!\:\!\plus\:\!00$\\
28&4418156892309715456&$0.0184$&$-1.5275$&$227.7144$&$-2.7810$&$-2.9716$&$0.9151$&$18.6426$&$1.2555\text{E}\:\!\:\!\plus\:\!01$\\
29&4418142117622280192&$-0.2247$&$-1.6319$&$227.7547$&$-2.2909$&$-2.6730$&$0.5875$&$17.3171$&$8.1503\text{E}\:\!\:\!\plus\:\!00$\\
30&4418143968756451968&$0.1305$&$-1.5872$&$227.8667$&$-2.7425$&$-2.7935$&$0.9761$&$19.4101$&$5.1128\text{E}\:\!\:\!\plus\:\!00$\\
31&4418261930029664256&$0.1726$&$-1.2314$&$227.9615$&$-2.4465$&$-2.4811$&$0.8722$&$17.4131$&$3.7794\text{E}\:\!\:\!\plus\:\!02$\\
32&4418265022406118784&$0.1443$&$-1.2007$&$227.9690$&$-2.5811$&$-2.7873$&$1.1245$&$17.1479$&$1.0070\text{E}\:\!\:\!\plus\:\!01$\\
33&4418300339922183040&$-0.2577$&$-0.9056$&$228.2730$&$-1.2246$&$-2.4189$&$1.1503$&$19.2080$&$5.7784\text{E}\:\!\:\!\plus\:\!00$\\
34&4418296800869119488&$-0.0699$&$-0.9458$&$228.3485$&$-2.5452$&$-2.5164$&$1.0382$&$18.2234$&$9.5543\text{E}\:\!\:\!\plus\:\!01$\\
35&4418305219004364416&$0.0308$&$-0.8191$&$228.4767$&$-2.3525$&$-3.1096$&$1.0936$&$18.2519$&$2.0354\text{E}\:\!\:\!\plus\:\!01$\\
36&4418876999410407808&$0.1810$&$-0.5154$&$228.4965$&$-2.6434$&$-2.8694$&$1.0682$&$18.4087$&$5.3204\text{E}\:\!\:\!\plus\:\!00$\\
37&4418292299742718336&$0.2783$&$-0.8791$&$228.5547$&$-3.1739$&$-2.0598$&$1.2035$&$18.9636$&$4.8350\text{E}\:\!\:\!\plus\:\!00$\\
38&4418688536245813888&$0.0397$&$-0.5523$&$228.5716$&$-2.8887$&$-2.5161$&$1.0884$&$17.9491$&$8.2213\text{E}\:\!\:\!\plus\:\!00$\\
39&4418306490314700416&$-0.0735$&$-0.7719$&$228.5788$&$-2.9618$&$-2.6038$&$1.0582$&$18.5771$&$1.1654\text{E}\:\!\:\!\plus\:\!01$\\
40&4418306387235478784&$0.0103$&$-0.7888$&$228.5841$&$-2.9801$&$-2.6432$&$1.1172$&$18.7119$&$1.0503\text{E}\:\!\:\!\plus\:\!01$\\
41&4418306490314700288&$0.0571$&$-0.7728$&$228.5863$&$-2.6047$&$-2.1894$&$1.1371$&$17.5508$&$1.8540\text{E}\:\!\:\!\plus\:\!02$\\
42&4418307899063987584&$0.4335$&$-0.7168$&$228.6229$&$-1.9881$&$-2.0470$&$0.9793$&$19.0673$&$1.4992\text{E}\:\!\:\!\plus\:\!01$\\
43&4418687161856244096&$-0.0988$&$-0.6207$&$228.6410$&$-2.4472$&$-2.8298$&$0.7099$&$17.1817$&$4.8383\text{E}\:\!\:\!\plus\:\!00$\\
44&4418889815592901248&$0.0445$&$-0.2970$&$228.6474$&$-2.2003$&$-2.2111$&$1.0924$&$17.7308$&$1.2297\text{E}\:\!\:\!\plus\:\!02$\\
45&4418683485364220160&$0.4711$&$-0.6612$&$228.6626$&$-2.3617$&$-3.3018$&$1.0654$&$19.4714$&$5.8224\text{E}\:\!\:\!\plus\:\!00$\\
46&4418696404625960448&$-0.5780$&$-0.4148$&$228.8240$&$-2.7140$&$-1.9526$&$0.7559$&$19.6628$&$6.0292\text{E}\:\!\:\!\plus\:\!00$\\
47&4418692040939128832&$0.2891$&$-0.5483$&$228.8315$&$-2.1215$&$-3.1551$&$1.1364$&$19.1036$&$8.0271\text{E}\:\!\:\!\plus\:\!00$\\
48&4418698397490799232&$-0.3326$&$-0.3732$&$228.9090$&$-2.4365$&$-1.7305$&$0.6510$&$20.0514$&$6.3760\text{E}\:\!\:\!\plus\:\!00$\\
49&4418679396555391616&$0.2885$&$-0.5816$&$228.9582$&$-2.5619$&$-2.6261$&$1.1009$&$18.5081$&$5.8196\text{E}\:\!\:\!\plus\:\!00$\\
50&4418723583179057024&$0.1416$&$-0.2626$&$228.9601$&$-2.5273$&$-2.7952$&$1.1119$&$17.4493$&$1.2010\text{E}\:\!\:\!\plus\:\!01$\\
51&4418723583179058944&$-0.0373$&$-0.2577$&$228.9661$&$-2.1354$&$-2.8744$&$1.0769$&$19.5808$&$9.7810\text{E}\:\!\:\!\plus\:\!00$\\
52&4418693724566344192&$0.3241$&$-0.4672$&$229.0063$&$-2.7870$&$-2.5274$&$1.1007$&$18.3871$&$4.6609\text{E}\:\!\:\!\plus\:\!00$\\
53&4418926855391616000&$-0.3033$&$+0.0552$&$229.0700$&$-2.0931$&$-2.7288$&$1.0322$&$19.2253$&$5.8788\text{E}\:\!\:\!\plus\:\!00$\\
54&4418724923208864128&$0.3599$&$-0.2139$&$229.0981$&$-3.3158$&$-2.1848$&$0.9962$&$19.1438$&$5.1160\text{E}\:\!\:\!\plus\:\!00$\\
55&4418734165978521728&$-0.0129$&$-0.0736$&$229.1446$&$-2.4978$&$-2.5807$&$0.4412$&$17.4019$&$5.0599\text{E}\:\!\:\!\plus\:\!01$\\
56&4418727225312005504&$0.1660$&$-0.2543$&$229.1711$&$-2.4135$&$-2.2349$&$1.0770$&$18.4655$&$2.0530\text{E}\:\!\:\!\plus\:\!01$\\
57&4418726404973230080&$0.1299$&$-0.2695$&$229.2028$&$-2.3950$&$-2.1456$&$0.8950$&$17.5034$&$7.0368\text{E}\:\!\:\!\plus\:\!01$\\
58&4419023234457620352&$-0.0293$&$+0.2043$&$229.2407$&$-2.4698$&$-2.3608$&$1.0275$&$17.9991$&$1.4207\text{E}\:\!\:\!\plus\:\!02$\\
59&4419073575769649664&$0.5628$&$+0.5239$&$229.3062$&$-1.9015$&$-2.2912$&$1.0387$&$18.1476$&$7.4501\text{E}\:\!\:\!\plus\:\!00$\\
60&4419026842229420800&$-0.1750$&$+0.2297$&$229.3883$&$-2.7700$&$-2.0247$&$1.1586$&$18.8187$&$1.6267\text{E}\:\!\:\!\plus\:\!01$\\
61&4419052405874782592&$0.2759$&$+0.3638$&$229.5968$&$-1.8983$&$-2.9869$&$1.1136$&$18.5392$&$4.7960\text{E}\:\!\:\!\plus\:\!00$\\
62&4419078145614843776&$0.4212$&$+0.6351$&$229.6086$&$-2.3773$&$-2.0128$&$1.1781$&$19.0422$&$1.5265\text{E}\:\!\:\!\plus\:\!01$\\
63&4419068108275272448&$-0.2572$&$+0.6591$&$229.7104$&$-1.4688$&$-2.4685$&$0.9486$&$19.7550$&$6.3516\text{E}\:\!\:\!\plus\:\!00$\\

\bottomrule

\end{tabular}
\end{center}

\label{sel0}
\end{table*}

\begin{table*}\addtocounter{table}{-1}
\caption[]{\small{\textit{- continued}}}
\begin{center}
\begin{tabular}{rrrrrrrrrr}
\toprule

\multicolumn{1}{c}{N}
&\multicolumn{1}{c}{source\_id}
&\multicolumn{1}{c}{$\pi$}
&\multicolumn{1}{c}{$\delta$}
&\multicolumn{1}{c}{$\alpha$}
&\multicolumn{1}{c}{$\mu_\delta$}
&\multicolumn{1}{c}{$\mu_{\alpha*}$}
&\multicolumn{1}{c}{\scalebox{0.8}{$G_{\rm BP}\!-\!G_{\rm RP}$}}
&\multicolumn{1}{c}{$G$}
&\multicolumn{1}{c}{$\chi_{\rm sel}$}\\
&
&\multicolumn{1}{c}{\units{mas}}
&\multicolumn{1}{c}{\units{deg}}
&\multicolumn{1}{c}{\units{deg}}
&\multicolumn{1}{c}{\units{mas yr$^{-1}$}}
&\multicolumn{1}{c}{\units{mas yr$^{-1}$}}
&\multicolumn{1}{c}{\units{mag}}
&\multicolumn{1}{c}{\units{mag}}
&\multicolumn{1}{c}{\units{\scalebox{0.8}{yr$^{3}$ deg$^{-2}$ pc$^{-1}$ mas$^{-3}$}}}\\

\midrule

64&4418864767344458112&$-0.0649$&$+0.3228$&$229.7453$&$-2.1466$&$-2.5965$&$1.1113$&$17.9590$&$7.6772\text{E}\:\!\:\!\plus\:\!00$\\
65&4420584403529433728&$0.1415$&$+0.8509$&$229.8402$&$-2.4470$&$-1.9695$&$1.0419$&$18.6502$&$2.9890\text{E}\:\!\:\!\plus\:\!01$\\
66&4420577943898627968&$-0.5387$&$+0.8701$&$230.0352$&$-2.1109$&$-2.9124$&$0.9411$&$19.3734$&$5.3310\text{E}\:\!\:\!\plus\:\!00$\\
67&4420385323205469568&$-0.2093$&$+0.7329$&$230.1499$&$-2.6252$&$-2.1064$&$0.9883$&$18.7577$&$5.4691\text{E}\:\!\:\!\plus\:\!00$\\
68&4420607768151633408&$-0.1165$&$+1.1328$&$230.2065$&$-2.3018$&$-2.7923$&$1.0686$&$17.8563$&$2.3135\text{E}\:\!\:\!\plus\:\!01$\\
69&4420603301385612928&$-0.3072$&$+1.0489$&$230.3098$&$-2.7339$&$-2.0564$&$1.0515$&$18.1806$&$2.5285\text{E}\:\!\:\!\plus\:\!01$\\
70&4420616289367861760&$0.1482$&$+1.2596$&$230.4004$&$-2.1897$&$-2.6394$&$1.1677$&$18.6286$&$5.1850\text{E}\:\!\:\!\plus\:\!01$\\
71&4420608562721387776&$0.2343$&$+1.0585$&$230.4380$&$-2.4493$&$-2.2273$&$1.0231$&$18.2679$&$5.9114\text{E}\:\!\:\!\plus\:\!01$\\
72&4420616048849698304&$0.1699$&$+1.2814$&$230.4905$&$-2.0212$&$-1.9596$&$1.0675$&$18.3503$&$1.0720\text{E}\:\!\:\!\plus\:\!02$\\
73&4420708888861661440&$-0.3825$&$+1.3479$&$230.6310$&$-1.9221$&$-2.4611$&$0.9986$&$18.0155$&$2.2752\text{E}\:\!\:\!\plus\:\!01$\\
74&4420717139494655360&$-0.3145$&$+1.4909$&$230.9006$&$-2.8521$&$-2.1866$&$1.0547$&$17.9769$&$8.1542\text{E}\:\!\:\!\plus\:\!00$\\
75&4420528568955338112&$0.3247$&$+1.4158$&$230.9018$&$-1.9665$&$-3.1172$&$0.9347$&$18.5703$&$1.3916\text{E}\:\!\:\!\plus\:\!01$\\
76&4420553479766045824&$0.1958$&$+1.5626$&$231.0663$&$-1.8110$&$-3.3450$&$0.9903$&$19.1601$&$1.2707\text{E}\:\!\:\!\plus\:\!01$\\
77&4420744283688076672&$0.3469$&$+1.6499$&$231.0725$&$-2.8218$&$-1.8201$&$1.0634$&$18.6081$&$1.0371\text{E}\:\!\:\!\plus\:\!01$\\
78&4420749364634207104&$0.4852$&$+1.6694$&$231.1743$&$-2.4018$&$-3.7009$&$1.2444$&$19.8641$&$5.1075\text{E}\:\!\:\!\plus\:\!00$\\
79&4421127944526566272&$-0.2777$&$+1.8523$&$231.3147$&$-3.3017$&$-2.4256$&$0.8227$&$19.4204$&$5.1616\text{E}\:\!\:\!\plus\:\!00$\\
80&4421128077670137472&$-0.3001$&$+1.8510$&$231.3650$&$-2.1888$&$-3.1238$&$1.0839$&$19.0124$&$6.0021\text{E}\:\!\:\!\plus\:\!00$\\
81&4420939958103532416&$0.2914$&$+1.8132$&$231.3706$&$-2.4673$&$-1.3992$&$1.0720$&$18.7013$&$4.9887\text{E}\:\!\:\!\plus\:\!00$\\
82&4420942432004721408&$-0.5258$&$+1.9198$&$231.5243$&$-1.6142$&$-2.4104$&$0.9329$&$18.6265$&$2.0817\text{E}\:\!\:\!\plus\:\!01$\\
83&4420974111684168320&$0.7073$&$+2.0716$&$231.8210$&$-2.9927$&$-2.4544$&$1.0667$&$19.6251$&$5.2512\text{E}\:\!\:\!\plus\:\!00$\\
84&4420970057233967360&$-0.0405$&$+2.0212$&$231.8757$&$-2.6548$&$-2.9159$&$1.0609$&$18.2706$&$5.5142\text{E}\:\!\:\!\plus\:\!00$\\
85&4420973256984758912&$0.3024$&$+2.1114$&$231.9175$&$-3.2737$&$-1.8754$&$1.1691$&$19.7097$&$5.4066\text{E}\:\!\:\!\plus\:\!00$\\
86&4420985003720258688&$0.1753$&$+2.2946$&$232.0258$&$-1.5876$&$-3.0622$&$0.8244$&$20.1691$&$5.0991\text{E}\:\!\:\!\plus\:\!00$\\
87&4420985553476080768&$0.1695$&$+2.3329$&$232.0740$&$-2.3067$&$-2.7102$&$0.6363$&$19.7382$&$1.4372\text{E}\:\!\:\!\plus\:\!01$\\
88&4421270739303308032&$0.3234$&$+2.4940$&$232.0890$&$-1.7718$&$-2.4556$&$0.7487$&$17.3115$&$1.8315\text{E}\:\!\:\!\plus\:\!01$\\
89&4421075889522936832&$0.1513$&$+2.3026$&$232.1542$&$-1.2829$&$-2.2284$&$1.1461$&$18.8079$&$2.1412\text{E}\:\!\:\!\plus\:\!01$\\
90&4420980914911383552&$0.0622$&$+2.2497$&$232.1790$&$-1.1811$&$-1.6482$&$0.9222$&$19.7160$&$5.5746\text{E}\:\!\:\!\plus\:\!00$\\
91&4421074820076313088&$0.0669$&$+2.2627$&$232.2309$&$-2.7257$&$-2.5686$&$1.1896$&$19.8201$&$6.3718\text{E}\:\!\:\!\plus\:\!00$\\
92&4421075644710040960&$-0.1957$&$+2.3202$&$232.2445$&$-2.4874$&$-2.1497$&$1.0964$&$17.9998$&$8.3941\text{E}\:\!\:\!\plus\:\!01$\\
93&4420967892569667200&$0.0372$&$+2.1748$&$232.2447$&$-1.9914$&$-1.7972$&$1.0749$&$18.4407$&$2.0904\text{E}\:\!\:\!\plus\:\!01$\\
94&4421063034685042048&$0.0499$&$+2.2580$&$232.2667$&$-1.7235$&$-2.9694$&$1.1849$&$18.4620$&$1.1119\text{E}\:\!\:\!\plus\:\!01$\\
95&4421279741554833664&$-0.0775$&$+2.6654$&$232.3405$&$-2.6957$&$-1.9749$&$0.8748$&$19.0484$&$6.0000\text{E}\:\!\:\!\plus\:\!00$\\
96&4421086261868313216&$-0.3842$&$+2.5208$&$232.5209$&$-2.4458$&$-2.3528$&$1.0164$&$18.5634$&$1.0815\text{E}\:\!\:\!\plus\:\!01$\\
97&4421118388224664448&$-0.2525$&$+2.7875$&$232.5409$&$-1.8784$&$-2.6760$&$1.0045$&$18.8868$&$8.5105\text{E}\:\!\:\!\plus\:\!00$\\
98&4421120896484661504&$-0.0837$&$+2.7778$&$232.8360$&$-2.2849$&$-3.5261$&$1.2619$&$19.8009$&$4.7386\text{E}\:\!\:\!\plus\:\!00$\\
99&4421122648831334784&$0.3999$&$+2.8572$&$232.8419$&$-1.3443$&$-2.8624$&$1.1055$&$19.5430$&$7.7608\text{E}\:\!\:\!\plus\:\!00$\\
100&4421121102643098880&$-0.0570$&$+2.8128$&$232.8420$&$-3.1021$&$-2.8906$&$1.0554$&$19.5488$&$5.8235\text{E}\:\!\:\!\plus\:\!00$\\
101&4427109146047429120&$-0.0208$&$+2.7260$&$233.0561$&$-2.4877$&$-2.6084$&$1.0095$&$18.4947$&$2.9412\text{E}\:\!\:\!\plus\:\!01$\\
102&4421056922947507584&$-0.1575$&$+2.6563$&$233.0629$&$-1.7088$&$-2.6912$&$0.9611$&$17.5178$&$7.0492\text{E}\:\!\:\!\plus\:\!00$\\
103&4427115605678338048&$0.5960$&$+2.8795$&$233.0941$&$-1.9663$&$-2.2967$&$1.0083$&$19.2071$&$1.3374\text{E}\:\!\:\!\plus\:\!01$\\
104&4427116365888010496&$-0.2454$&$+2.9255$&$233.1339$&$-1.8015$&$-1.8290$&$0.8261$&$19.4928$&$1.3035\text{E}\:\!\:\!\plus\:\!01$\\
105&4427116091010101888&$-0.3001$&$+2.9078$&$233.1520$&$-1.6010$&$-2.4964$&$1.0085$&$18.9371$&$2.0755\text{E}\:\!\:\!\plus\:\!01$\\
106&4427119350890013312&$-0.2800$&$+2.9388$&$233.2521$&$-2.0540$&$-1.7483$&$1.0091$&$18.9006$&$9.9676\text{E}\:\!\:\!\plus\:\!00$\\
107&4427149587458915328&$0.2883$&$+3.1325$&$233.3049$&$-1.9897$&$-2.8823$&$0.9629$&$19.6203$&$6.1447\text{E}\:\!\:\!\plus\:\!00$\\
108&4427072385423018112&$-0.0852$&$+2.9171$&$233.4837$&$-1.3629$&$-2.5185$&$0.8760$&$19.2556$&$4.8170\text{E}\:\!\:\!\plus\:\!00$\\
109&4427159070746889088&$0.0524$&$+3.3349$&$233.5535$&$-2.0658$&$-2.0261$&$1.0945$&$18.0417$&$1.4151\text{E}\:\!\:\!\plus\:\!02$\\
110&4427267342578834816&$-0.1346$&$+3.4730$&$233.6187$&$-2.3678$&$-2.0626$&$1.1503$&$17.8807$&$1.5292\text{E}\:\!\:\!\plus\:\!01$\\
111&4427108080895293440&$-0.2145$&$+3.2197$&$233.7217$&$-2.6400$&$-1.9216$&$0.8551$&$19.9234$&$7.9206\text{E}\:\!\:\!\plus\:\!00$\\
112&4427252391796247936&$-0.4364$&$+3.4275$&$233.9223$&$-2.3541$&$-1.6143$&$1.0031$&$18.5629$&$8.8254\text{E}\:\!\:\!\plus\:\!00$\\
113&4427286068636375552&$0.1463$&$+3.6697$&$234.1154$&$-1.8386$&$-2.5410$&$0.8207$&$17.4965$&$4.3439\text{E}\:\!\:\!\plus\:\!01$\\
114&4427242118235062400&$0.3643$&$+3.6487$&$234.4303$&$-1.8013$&$-2.5729$&$1.2051$&$18.6923$&$3.2961\text{E}\:\!\:\!\plus\:\!01$\\
115&4427620281515156608&$0.0090$&$+3.8090$&$234.6403$&$-1.8909$&$-2.7907$&$1.0151$&$18.9503$&$2.9022\text{E}\:\!\:\!\plus\:\!01$\\
116&4427616364504699264&$0.5170$&$+3.9037$&$234.9541$&$-2.2568$&$-1.8675$&$1.1695$&$18.9193$&$7.3622\text{E}\:\!\:\!\plus\:\!00$\\
117&4427617051699467264&$-0.3315$&$+3.9052$&$234.9959$&$-1.6459$&$-2.7952$&$0.9918$&$19.0932$&$8.2553\text{E}\:\!\:\!\plus\:\!00$\\
118&4427641206595577728&$0.1592$&$+4.0049$&$235.1233$&$-3.0101$&$-2.6033$&$0.9046$&$19.6347$&$4.7413\text{E}\:\!\:\!\plus\:\!00$\\
119&4427638079859372288&$-0.0039$&$+3.9687$&$235.1463$&$-1.8371$&$-1.1354$&$0.9195$&$19.4172$&$4.6017\text{E}\:\!\:\!\plus\:\!00$\\
120&4427657149514263680&$-0.0730$&$+4.1916$&$235.1697$&$-2.0171$&$-3.0982$&$1.1434$&$19.2056$&$6.1437\text{E}\:\!\:\!\plus\:\!00$\\
121&4427749061814709504&$0.2200$&$+4.3595$&$235.5140$&$-1.6084$&$-3.0522$&$0.9691$&$19.2026$&$4.7347\text{E}\:\!\:\!\plus\:\!00$\\
122&4427746076813323520&$-0.0727$&$+4.3656$&$235.6363$&$-2.5033$&$-1.9675$&$1.1675$&$18.5294$&$1.7471\text{E}\:\!\:\!\plus\:\!01$\\
123&4424743718578674688&$-0.0442$&$+4.2818$&$235.9371$&$-2.4113$&$-2.2675$&$1.0670$&$17.3253$&$1.0274\text{E}\:\!\:\!\plus\:\!01$\\
124&4424779727584949120&$0.1786$&$+4.6100$&$236.2725$&$-1.7183$&$-2.3518$&$1.0645$&$18.5830$&$2.3745\text{E}\:\!\:\!\plus\:\!01$\\
125&4426303478901609344&$0.0403$&$+4.9487$&$237.0251$&$-2.1590$&$-1.7255$&$1.1713$&$18.9585$&$6.5246\text{E}\:\!\:\!\plus\:\!00$\\
126&4426315848407518336&$0.3195$&$+5.1591$&$237.1373$&$-1.8922$&$-1.9475$&$1.0778$&$18.8793$&$5.0225\text{E}\:\!\:\!\plus\:\!00$\\

\bottomrule
\end{tabular}
\end{center}

\label{sel1}
\end{table*}

\begin{table}
\caption[]{\small{Stars from Table \ref{sel0} with radial velocity measured by \citet{2017ApJ...842..120I}.}}
\begin{center}
\begin{tabular}{rrrr}
\toprule
\multicolumn{1}{c}{N}&\multicolumn{1}{c}{source\_id}&\multicolumn{1}{c}{$v_r$}&\multicolumn{1}{c}{$\epsilon_{v_r}$}\\
&&\multicolumn{1}{c}{\units{km s$^{-1}$}}&\multicolumn{1}{c}{\units{km s$^{-1}$}}\\
\midrule
28&4418156892309715456&-58.64&2.00\\
32&4418265022406118784&-56.36&1.64\\
35&4418305219004364416&-69.85&3.90\\
38&4418688536245813888&-54.24&2.19\\
41&4418306490314700288&-54.12&1.30\\
54&4418724923208864128&-60.28&1.54\\
68&4420607768151633408&-48.15&1.75\\
74&4420717139494655360&-48.75&2.50\\
76&4420553479766045824&-62.16&0.98\\
80&4421128077670137472&-47.55&3.94\\
82&4420942432004721408&-53.68&2.91\\
84&4420970057233967360&-58.02&1.46\\
92&4421075644710040960&-49.59&1.65\\
105&4427116091010101888&-33.85&2.97\\
106&4427119350890013312&-54.28&3.80\\
\bottomrule
\end{tabular}
\end{center}
\label{rv_table}
\end{table}


\section{Definition of the likelihood function}\label{A}

The likelihood function is composed of the product of the likelihoods corresponding to the constraints enumerated in Section \ref{SM} plus the stellar steams, all together denoted by $d$:
\begin{equation}
\mathcal{L}\var{d|\theta} \equiv \mathcal{L}_{c}\var{d_{\rm c}|\theta} \, \mathcal{L}_{\rm str}\var{d_{\rm str}|\theta} ~.
\end{equation}

We assume that the model of the constraints $d_{\rm c} \equiv \big\{ f_\varSigma, M_{\rm b}, M_{200}, \mu_{l}, K_z, V_{\rm c}^{1}, \ldots, V_{\rm c}^{38} \big\}$ is a Dirac's delta distribution and each observational measurement a Gaussian distribution with mean $\mu$ and standard deviation $\sigma$. For the constraint $i$ we have:
\begin{equation}
\int_{-\infty}^{\infty} \delta\vvar{x-d_{\rm c}^i\var{\theta}} \, G\vvar{x|\mu_i,\sigma_i^2} \diff x = G\vvar{d_{\rm c}^i\var{\theta}|\mu_i,\sigma_i^2} ~,
\end{equation}
from which we get the likelihood function: 
\begin{equation}
\mathcal{L}_{d}\var{d_{\rm c}|\theta} = \prod_{i=1}^{43} G\vvar{d_{\rm c}^i\var{\theta}|\mu_i,\sigma_i^2} ~.
\end{equation}

The likelihood function of a stellar stream is defined from a phase-space probability density model of the stream. We define this model in Heliocentric spherical coordinates because we have the observed stars in this coordinate system. This model is constructed from a numerical simulation using a Kernel Density Estimation method introduced in \citetalias{2019MNRAS.488.1535P}. This method is based on a Gaussian kernel, where the covariance matrix of each Gaussian is calculated with the neighbouring points, weighting their contribution according to an inverse function of the distance between points. In this way, basically taking into account the nearest neighbours, the kernels are optimized for the characteristics of each section of the stream. This is specially required at the extremes of the stream, where there are a few points and are separated by large distances.

Given a simulation of the stellar stream made of $N$ stars, we locate the mean of a Gaussian distribution at the phase-space position $\eta^\nu_n$ of each $n$ star, and we compute its covariance matrix $\varXi^{\nu\epsilon}_n$ from the position of the neighbouring stars:
\begin{equation}
\varXi_n^{\nu\epsilon} = \left(\, \sum^N_{m=1} c_{nm} \! \right)^{-1} \sum^N_{m=1} c_{nm} \:\! (\eta^\nu_m - \eta^\nu_n) \:\! (\eta^\epsilon_m - \eta^\epsilon_n) ~,
\end{equation}
where the indices $\nu, \epsilon \equiv (\pi, \delta, \alpha, v_r, \mu_\delta, \mu_\alpha)$. The weighting factors determine the kernel size, and are defined as:
\begin{equation}
c_{nm} = (l_{\rm 0}+l_{nm})^{-9/2} ~,\qquad\, l_{nm}^{\:\!2} = \sum_{j=1}^3 (x_m^k-x_n^k)^2 ~,
\end{equation}
where $x^k$ are the Galactocentric Cartesian coordinates of each star at present time. The constant $l_{\rm 0} = 250$ pc and the slope $9/2$ have been optimised in \citetalias{2019MNRAS.488.1535P} to reproduce properly the distribution of the escaped stars.

\begin{figure}
\includegraphics[width=1.0\columnwidth]{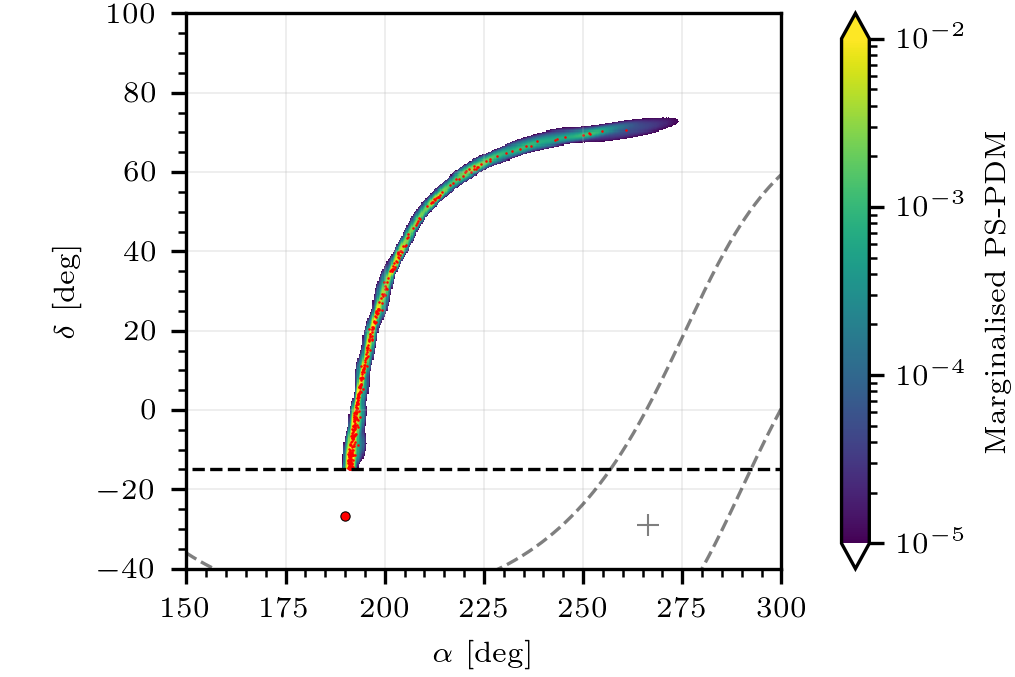}\\
\includegraphics[width=1.0\columnwidth]{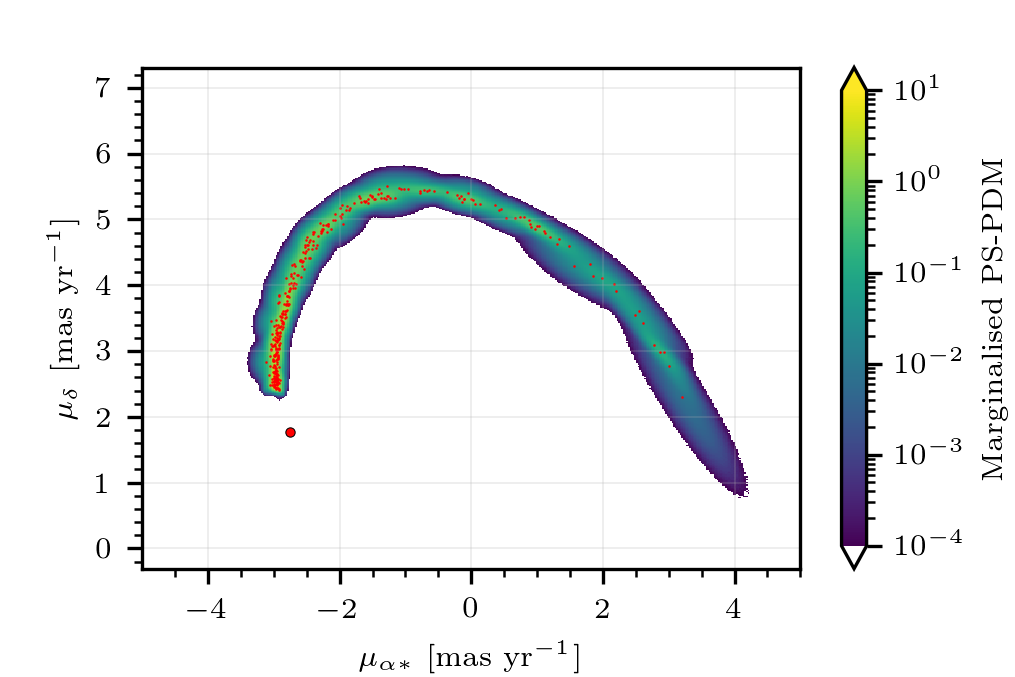}
\caption{Marginalised phase-space probability density model of the M68 stream (PS-PDM), computed for the best-fitting parameters of the model with all streams together. The small red dots mark the positions of the simulated stars used to compute the density model, and the large red dot marks the position of the cluster. \textit{Top panel:} Sky coordinates space. The horizontal black dashed line marks the $\delta=-15$ deg limit used to define the density model. The grey dashed lines mark the Milky Way disc at $b = \pm15$ deg and the grey cross marks the Galactic centre. \textit{Bottom panel:} Proper motions space.}
\label{denisty_model}
\end{figure}

To show the typical kernel size obtained using this method, we plot in Figure \ref{denisty_model} as an example, the resulting marginalised phase space density model of the M68 stream for sky coordinates and proper motions. This model is computed using the best-fitting parameters obtained with all the streams together, which we list in Table \ref{table_res1}. We also mark the position of the simulated stars used to construct the model with small red dots and the cluster with a large red dot. The dashed horizontal line in the top panel marks the $\delta=-15$ deg limit that we use to define the density model, which slightly exceeds the limit of the observational data, which we set at $\delta=-8$ deg (Section \ref{str_m68}). We note that the protuberances in the density profile visible along the edge defined by the lower limit of the log-scale are the result of a few stars deviating from the mean track. This is clearly visible in proper motion space within the interval of about $\mu_{\alpha *} \in \Range{-2}{0}$ mas yr$^{-1}$ and at the far end of the distribution away from the cluster. The central zone of the distribution, which is more than 4 orders of magnitude denser than the edge, is more homogeneous. Such irregularities depend on the particular random sample of simulated stars used to compute the density model. The total number of simulated stars is the result of a compromise between a smooth model with many stars and a fast evaluation of the posterior function, which requires a minimum number of stars. For NGC 3201, M68, and Palomar 5, we use 240, 300, and 170 simulated stars, respectively, along the section of the stream where we have observational data. We find that these numbers are sufficient to build a smooth density model and evaluate the likelihood function with minimal computational time.

Next, we assume that the observed stars $d_{\rm str}$ follow a Gaussian distribution centred at the mean phase-space position of the star $w^\nu$, where the covariance matrix $\varSigma^{\nu\epsilon}$ is the value of the observational errors and their correlations. If the stars do not have radial velocity we take $v_r=0 \pm 10^3$  ${\rm km\,s}^{-1}$. This is a value with an uncertainty much bigger than the expected distribution of radial velocities of the stellar stream. It is almost equivalent to use a uniform distribution for the missing radial velocity, but simplifies the definition of the likelihood function. For the $j$ observed star we have:
\begin{equation}
\begin{split}
\sum_{n=1}^{N} \, \int_{-\infty}^{\infty} G\vvar{w^\nu|\eta_n^\nu,\varXi_n^{\nu\epsilon}} \, G\vvar{w^\nu|\nu_j^\nu,\varSigma_j^{\nu\epsilon}} \diff^6 w = \\ = \sum_{n=1}^{N} G\vvar{\eta^\nu_n|\nu_j^\nu,\varXi_n^{\nu\epsilon}+\varSigma_j^{\nu\epsilon}} ~.
\end{split}
\end{equation}
If $J$ is the total number of observed stars in the stream, we get the following likelihood function:
\begin{equation}
\mathcal{L}_{\rm str}\var{d_{\rm str}|\theta} = \prod_{j=1}^{J} \sum_{n=1}^{N} G\vvar{\eta^\nu_n|\nu_j^\nu,\varXi_n^{\nu\epsilon}+\varSigma_j^{\nu\epsilon}} ~.
\end{equation}


\section{Stream coordinates}\label{B}

To minimise the time required to evaluate the likelihood function, we apply a method to obtain an approximate distribution of stream stars from a pre-calculated simulation. 

To carry out this simulation, we choose the mean position of the globular cluster and a fiducial potential of the Galaxy. We take the orbit of a globular cluster in Galactocentric Cartesian coordinates $x^i_{\rm o}$, where $i=(x,y,z)$, during $T$ Myr backwards and forwards from the present position of the cluster. We also take the current position of a simulated stream stars $x^i_{e}$. For each star $e$, we compute the closest point of the orbit to the star using an Euclidean distance. This point is expressed in function of the parameter $t$:
\begin{equation}
\hat{t}_e \equiv \operatorname*{argmin}_{t\in[-T,T]}\;\!\!\!\vvar{d_e\var{t}} ~,\qquad\, d^{\:\!2}_e\var{t} = \sum_{i=1}^3 (x^i_{\rm o}\var{t}-x_e^i)^2 ~.
\end{equation}
Defining $v\equiv v^i_{\rm o}\var{\hat{t}_e}$ as the velocity of the cluster and
\begin{equation}
a\equiv \dfrac{\diff}{\diff t} \dfrac{v}{|v|} ~,
\end{equation}
and assuming $|v|>0$, $|a|>0$, and $|v \times a|>0$ for any $t$, we define the corresponding Frenet-Serret trihedron at the point $x^i_{\rm o}\var{\hat{t}_e}$ as:
\begin{equation}
e_1 = \frac{v}{|v|} ~,\qquad e_2 = \frac{a}{|a|} ~,\qquad e_3 = \frac{e_1 \times e_2}{|e_1 \times e_2|} ~.
\end{equation}
We store the parameter $\hat{t}_e$ and the position and velocity of the stream star expressed in the coordinate system defined by the vector basis $(e_1, e_2, e_3)$. 

When we evaluate the likelihood function for different values of the free parameters, we compute a new orbit of the cluster $\bar{x}_{\rm o}^i$. We assume that the stored values are independent of the orbit for small variations with respect to $x^i_{\rm o}$. Then, for each star $e$, we compute the Frenet-Serret trihedron corresponding to the position $\bar{x}_{\rm o}^i\var{\hat{t}_e}$, and we locate the star at the stored values in the reference frame defined by the new trihedron $(\bar{e}_1, \bar{e}_2, \bar{e}_3)$. Finally, we put back the stars on the Galactocentric Cartesian coordinate system to get an approximation of the stellar stream for the new values of the free parameters.

This method is based on the assumption that the internal structure of the stream and the offset between the stream and the cluster orbit does not change for small variations of the Galactic potential. We note that in action-angle coordinates, the misalignment angle between the stream and the cluster orbit is constant along the stream in the angle space \citep{2011MNRAS.413.1852E}. For a small variation of the potential, for example a variation $\Delta q^{\rm h}_\rho \Approx 15$ per cent, we expect variations of the cluster radial action of $\Delta J_R \Approx 20$ per cent and of the vertical action $\Delta J_z \Approx 10$ per cent. Such variations of the actions will result in a negligible change in the misalignment angle, as well as in the length and width of the stream. This is exemplified by a numerical calculation for a realistic model of the Milky Way \citep{2011MNRAS.414.2446M} and a GD-1-like orbit in Figure 2 of \citet{2013MNRAS.433.1813S}. When the stream is mapped to real space from the action-angle space, the constant misalignment angle translates into a constant offset between the stream and the cluster orbit.


\section{Halo flattening}\label{D}

In Table \ref{table_res_q}, we show the median with the $1\sigma$ levels and the mean with the standard deviation of the marginalised posterior probability density function of the axis ratio $q_\rho^{\rm h}$. We also include the best-fitting parameters of a log-Normal probability density function defined as:
\begin{equation}
\logN\!\var{x|\mu,\tau,\epsilon} \equiv \frac{1}{\sqrt{2\pi}(x-\mu)\tau} \exp\!\left[\frac{-\log\!\left(\frac{x-\mu}{\epsilon}\right)^2}{2\tau^2} \right] \,\,.
\end{equation}

\begin{table}
\caption[]{\small{Median with $1\sigma$ levels, mean and stardard deviation, and best-fitting parameters of a log-Normal distribution of the marginalised posterior probability density function of the flattening parameter $q_\rho^{\rm h}$. }}

\begin{center}
\begin{tabular}{lcll}
\toprule
&Median$^{+\sigma}_{-\sigma}$&${\rm Mean}\,\,\,\,\,s$&$\hspace{1.9em}\mu\hspace{1.85em}\tau\hspace{2.0em}\epsilon$\\
\midrule
\textbf{NGC 3201}&$2.06_{-0.86}^{+1.01}$&$2.13\,\,\,\,0.88$&\phantom{0}$-2.79\,\,\,\,0.18\,\,\,\,4.84$\\[4.5pt]
\textbf{M68}&$1.14_{-0.14}^{+0.21}$&$1.17\,\,\,\,0.18$&\phantom{+}\phantom{0}$0.60\,\,\,\,0.31\,\,\,\,0.54$\\[4.5pt]
\textbf{Palomar 5}&$1.01_{-0.09}^{+0.09}$&$1.01\,\,\,\,0.09$&\phantom{0}$-8.18\,\,\,\,0.01\,\,\,\,9.19$\\[4.5pt]
\textbf{All}&$1.06_{-0.05}^{+0.06}$&$1.07\,\,\,\,0.07$&\phantom{+}\phantom{0}$0.66\,\,\,\,0.16\,\,\,\,0.40$\\
\bottomrule
\end{tabular}
\end{center}

\label{table_res_q}
\end{table}


\section{Numerical results}\label{C}

Table \ref{table_res1} lists the median and $1\sigma$ levels of the marginalised free parameters and derived properties of the Galaxy potential
model. They are computed for each stream separately and for all streams together.
Asymmetric errors are given when the difference between the upper and lower uncertainty is larger than 20 per cent.

\begin{table*}
\caption[]{\small{Gaussian priors, median and $1\sigma$ levels of the marginalised free parameters of our models. We include the value of the constraints described in Section \ref{MassModel} and \ref{KCons}, and several derived properties of the Milky Way model.}}

\begin{center}
\begin{tabular}{lllllll}
\toprule
\multicolumn{2}{l}{\textbf{Parameter}}&\textbf{Gaussian Prior}&\textbf{NGC 3201}&\textbf{M68}&\textbf{Palomar 5}&\textbf{All}\\
\midrule

$R_{\Sun}$&\units{kpc}&$8.178\pm0.026$&$8.17\pm0.02$&$8.17\pm0.03$&$8.14\pm0.03$&$8.15_{-0.03}^{+0.02}$\\[4.5pt]
$U_{\Sun}$&\units{km s$^{-1}$}&$11.1\pm1.25$&$9.61\pm1.01$&$9.89\pm0.97$&$12.47\pm1.27$&$10.58_{-0.42}^{+0.89}$\\[4.5pt]
$V_{\Sun}$&\units{km s$^{-1}$}&$12.24\pm2.05$&$14.58\pm1.43$&$15.24\pm1.56$&$13.39\pm1.57$&$13.64_{-1.2}^{+1.84}$\\[4.5pt]
$W_{\Sun}$&\units{km s$^{-1}$}&$7.25\pm0.62$&$7.61\pm0.56$&$7.45\pm0.59$&$7.22\pm0.64$&$7.33\pm0.51$\\
\midrule
$\rho_{0}^{\rm b}$&\units{$10^{10} \,$M$_{\Sun}$ kpc$^{-3}$}&&$9.80\pm1.02$&$9.62\pm1.03$&$9.97_{-1.08}^{+0.78}$&$9.84_{-0.95}^{+0.69}$\\
\midrule
$\varSigma_{\rm n}$&\units{$10^{9} \,$M$_{\Sun}$ kpc$^{-2}$}&&$0.92\pm0.22$&$1.19\pm0.13$&$0.9_{-0.15}^{+0.21}$&$1.25_{-0.2}^{+0.14}$\\[4.5pt]
$h_{\rm n}$&\units{kpc}&$2.6\pm0.5$&$3.01_{-0.23}^{+0.29}$&$2.88_{-0.14}^{+0.21}$&$2.99\pm0.25$&$2.78_{-0.1}^{+0.18}$\\[4.5pt]
$z_{\rm n}$&\units{kpc}&$0.3\pm0.05$&$0.31\pm0.05$&$0.30\pm0.05$&$0.31\pm0.05$&$0.31_{-0.03}^{+0.04}$\\
\midrule
$\varSigma_{\rm k}$&\units{$10^{8} \,$M$_{\Sun}$ kpc$^{-2}$}&&$4.17_{-1.66}^{+2.66}$&$4.75\pm2.06$&$4.25_{-1.7}^{+2.39}$&$3.77_{-1.37}^{+2.44}$\\[4.5pt]
$h_{\rm k}$&\units{kpc}&$2.0\pm0.2$&$1.97\pm0.19$&$1.98\pm0.18$&$1.93\pm0.18$&$2.06_{-0.21}^{+0.14}$\\[4.5pt]
$z_{\rm k}$&\units{kpc}&$0.9\pm0.18$&$0.93\pm0.16$&$0.90\pm0.18$&$0.91\pm0.17$&$0.79_{-0.12}^{+0.25}$\\
\midrule
$\rho_{0}^{\rm h}$&\units{$10^{7} \,$M$_{\Sun}$ kpc$^{-3}$}&&$1.92_{-1.6}^{+2.85}$&$2.94_{-1.44}^{+2.22}$&$2.0_{-1.5}^{+2.88}$&$1.84_{-0.62}^{+1.05}$\\[4.5pt]
$\alpha$&&&$0.68\pm0.64$&$-0.23\pm0.39$&$0.73\pm0.45$&$0.06\pm0.22$\\[4.5pt]
$a_{1}$&\units{kpc}&&$12.58_{-6.13}^{+20.06}$&$18.63_{-5.41}^{+10.08}$&$11.22_{-5.1}^{+17.14}$&$17.36_{-2.74}^{+9.77}$\\[4.5pt]
$\beta$&&&$3.19_{-0.61}^{+1.19}$&$3.73_{-0.56}^{+0.83}$&$2.77_{-0.33}^{+0.93}$&$3.29_{-0.28}^{+0.66}$\\[4.5pt]
$q_{\rho}^{\rm h}$&&&$2.06\pm0.93$&$1.14_{-0.14}^{+0.21}$&$1.01\pm0.09$&$1.06\pm0.06$\\
\midrule
$r_{\rm h}^{\SM{NGC3201}}$&\units{kpc}&$4.9\pm0.11$&$4.82\pm0.02$&&&$4.83\pm0.02$\\[4.5pt]
$v_r^{\SM{NGC3201}}$&\units{km s$^{-1}$}&$494.34\pm0.14$&$494.32\pm0.14$&&&$494.31\pm0.13$\\[4.5pt]
$\mu_\delta^{\SM{NGC3201}}$&\units{mas yr$^{-1}$}&$-1.991\pm0.044$&$-1.962\pm0.023$&&&$-1.931_{-0.019}^{+0.03}$\\[4.5pt]
$\mu_{\alpha*}^{\SM{NGC3201}}$&\units{mas yr$^{-1}$}&$8.324\pm0.044$&$8.309\pm0.042$&&&$8.293\pm0.056$\\
\midrule
$r_{\rm h}^{\SM{M68}}$&\units{kpc}&$10.3\pm0.52$&&$10.01_{-0.11}^{+0.08}$&&$10.03\pm0.06$\\[4.5pt]
$v_r^{\SM{M68}}$&\units{km s$^{-1}$}&$-92.99\pm0.22$&&$-92.95\pm0.22$&&$-92.9_{-0.29}^{+0.2}$\\[4.5pt]
$\mu_\delta^{\SM{M68}}$&\units{mas yr$^{-1}$}&$1.762\pm0.053$&&$1.766\pm0.027$&&$1.782\pm0.027$\\[4.5pt]
$\mu_{\alpha*}^{\SM{M68}}$&\units{mas yr$^{-1}$}&$-2.752\pm0.054$&&$-2.750\pm0.028$&&$-2.744_{-0.026}^{+0.02}$\\
\midrule
$r_{\rm h}^{\SM{Palomar\,5}}$&\units{kpc}&$20.6\pm0.2$&&&$21.19\pm0.15$&$21.20\pm0.15$\\[4.5pt]
$v_r^{\SM{Palomar\,5}}$&\units{km s$^{-1}$}&$-58.6\pm0.21$&&&$-58.44\pm0.20$&$-58.5_{-0.12}^{+0.17}$\\[4.5pt]
$\mu_\delta^{\SM{Palomar\,5}}$&\units{mas yr$^{-1}$}&$-2.646\pm0.064$&&&$-2.546\pm0.016$&$-2.544_{-0.018}^{+0.013}$\\[4.5pt]
$\mu_{\alpha*}^{\SM{Palomar\,5}}$&\units{mas yr$^{-1}$}&$-2.736\pm0.064$&&&$-2.533\pm0.017$&$-2.513_{-0.02}^{+0.015}$\\

\bottomrule

\end{tabular}
\end{center}

\label{table_res1}
\end{table*}

\begin{table*}\addtocounter{table}{-1}
\caption[]{\small{- \textit{continued}}}

\begin{center}
\begin{tabular}{lllllll}
\toprule
\multicolumn{2}{l}{\textbf{Parameter}}&\textbf{Gaussian Prior}&\textbf{NGC 3201}&\textbf{M68}&\textbf{Palomar 5}&\textbf{All}\\
\midrule
$f_\varSigma$&&$0.12\pm0.04$&$0.11\pm0.03$&$0.11\pm0.04$&$0.11\pm0.03$&$0.11\pm0.03$\\[4.5pt]
$K_z$&\units{$2\pi G$ M$_{\Sun}$ pc$^{-2}$}&$74\pm6$&$77.58\pm4.73$&$88.49\pm4.31$&$80.37\pm5.47$&$86.81_{-3.84}^{+2.93}$\\[4.5pt]
$ $&\units{km$^2$ pc$^{-1}$ s$^{-2}$}&$2\pm 0.16$&$2.10\pm0.13$&$2.39\pm0.12$&$2.17\pm0.15$&$2.33\pm0.09$\\
\midrule
$\mu_{l}$&\units{mas yr$^{-1}$}&$-6.379\pm0.026$&$-6.37\pm0.02$&$-6.35\pm0.02$&$-6.33\pm0.02$&$-6.32\pm0.02$\\[4.5pt]
$\Theta_0$&\units{km s$^{-1}$}&&$232.39\pm1.38$&$230.70\pm1.44$&$230.79\pm1.48$&$230.67\pm1.55$\\[4.5pt]
$\Theta_0 + V_{\Sun}$&\units{km s$^{-1}$}&&$246.95\pm1.14$&$245.92\pm1.14$&$244.15\pm1.12$&$244.38\pm0.91$\\
\midrule
$V_{\rm c}\,\units{R\,=\,5.27\,{\rm kpc}}$&\units{km s$^{-1}$}&$226.83\pm 7.07$&$230.74\pm 2.8$&$230.11\pm 2.92$&$229.49\pm 2.58$&$231.1_{-3.51}^{+2.32}$\\[4.5pt]
$V_{\rm c}\,\units{R\,=\,10.26\,{\rm kpc}}$&\units{km s$^{-1}$}&$225.68\pm 6.78$&$227.82\pm 1.28$&$225.38\pm 1.25$&$226.18\pm 1.28$&$224.85_{-1.32}^{+0.98}$\\[4.5pt]
$V_{\rm c}\,\units{R\,=\,15.22\,{\rm kpc}}$&\units{km s$^{-1}$}&$217.07\pm 6.58$&$214.55\pm 1.4$&$212.71\pm 1.35$&$213.33\pm 1.32$&$211.43_{-1.28}^{+1.64}$\\[4.5pt]
$V_{\rm c}\,\units{R\,=\,20.27\,{\rm kpc}}$&\units{km s$^{-1}$}&$199.84\pm 6.71$&$204.72\pm 1.92$&$205.66\pm 1.95$&$204.02\pm 1.96$&$204.88\pm2.2$\\[4.5pt]
$V_{\rm c}\,\units{R\,=\,24.82\,{\rm kpc}}$&\units{km s$^{-1}$}&$198.42\pm 8.67$&$198.76\pm 2.94$&$202.52\pm 2.84$&$198.54\pm 2.94$&$202.2_{-3.3}^{+2.45}$\\
\midrule
$M_{\rm b}$&\units{$10^{9} \,$M$_{\Sun}$}&$8.9\pm0.89$&$8.84\pm0.92$&$8.68\pm0.93$&$9.0_{-0.97}^{+0.71}$&$8.87_{-0.86}^{+0.62}$\\[4.5pt]
$M_{\rm d}^{\rm n}$&\units{$10^{10} \,$M$_{\Sun}$}&&$5.26\pm0.60$&$6.22\pm0.40$&$5.13\pm0.56$&$6.07\pm0.39$\\[4.5pt]
$M_{\rm d}^{\rm k}$&\units{$10^{10} \,$M$_{\Sun}$}&&$1.02\pm0.36$&$1.16\pm0.39$&$1.00\pm0.36$&$1.01_{-0.29}^{+0.4}$\\[4.5pt]
$M_{\Star}$&\units{$10^{10} \,$M$_{\Sun}$}&&$7.22\pm0.70$&$8.30\pm0.49$&$7.04\pm0.66$&$8.01\pm0.38$\\[4.5pt]
$M^{\rm h}_{200}$&\units{$10^{12} \,$M$_{\Sun}$}&&$0.95\pm0.23$&$0.94\pm0.22$&$0.94\pm0.24$&$1.08\pm0.22$\\[4.5pt]
$M_{200}$&\units{$10^{12} \,$M$_{\Sun}$}&&$1.03\pm0.23$&$1.03\pm0.23$&$1.02\pm0.25$&$1.18\pm0.23$\\
\midrule
$\rho_{\rm h}\var{R_{\Sun}}$&\units{$10^{6} \,$M$_{\Sun}$ kpc$^{-3}$}&&$5.2_{-0.77}^{+1.87}$&$5.66\pm0.84$&$7.43\pm0.86$&$5.95\pm0.60$\\[4.5pt]
$ $&\units{GeV cm$^{-3}$}&&$0.2_{-0.03}^{+0.07}$&$0.21\pm0.03$&$0.28\pm0.03$&$0.23\pm0.02$\\[4.5pt]
$r_{200}$&\units{kpc}&&$200.56\pm16.38$&$199.89_{-19.16}^{+13.52}$&$199.85\pm17.33$&$209.6_{-16.71}^{+12.79}$\\[4.5pt]
$c_{200}$&&&$13.48_{-3.06}^{+4.82}$&$8.24\pm0.58$&$10.43_{-1.92}^{+2.95}$&$7.86_{-0.44}^{+0.57}$\\
\bottomrule
\end{tabular}
\end{center}

\end{table*}



\bsp	
\label{lastpage}

\end{document}